\documentclass[english,twocolumn]{IEEEtran}
\IEEEoverridecommandlockouts
\usepackage[T1]{fontenc}
\usepackage[utf8]{inputenc}
\usepackage{color}
\usepackage[commentmarkup=footnote]{changes}
\definechangesauthor[name=Farzad, color=purple]{F}
\definecolor{documenT_Fontcolor}{rgb}{0, 0, 0}
\color{documenT_Fontcolor}
\usepackage{units}
\usepackage{mathtools}
\usepackage{nicefrac}
\usepackage{amsthm}
\usepackage{amsmath}
\usepackage{thmtools}
\usepackage{thm-restate}
\usepackage{hyperref}
\usepackage{amssymb}
\usepackage{graphicx}
\usepackage{esint}
\PassOptionsToPackage{normalem}{ulem}
\usepackage{ulem}
\usepackage{epstopdf}
\usepackage{tikz}
\usetikzlibrary{decorations.pathreplacing,intersections}
\usetikzlibrary{fadings,decorations.text}
\usepackage{color}
\usetikzlibrary{patterns}
\usetikzlibrary{positioning}
\usepackage{csquotes}
\presetkeys%
    {todonotes}%
    {inline}{}
    
\makeatletter

\usepackage[
backend=biber,
style=ieee,
doi=false,isbn=false,url=false,
sorting=nty
]{biblatex}
 
\theoremstyle{plain}
\newtheorem{thm}{\protect\theoremname}
\theoremstyle{plain}
\newtheorem{lem}[thm]{\protect\lemmaname}
\theoremstyle{definition}

\theoremstyle{plain}
\newtheorem{cor}{\protect\corollaryname}
\theoremstyle{definition}
\newtheorem{exmp}{\protect\examplename}
\usepackage{thmtools,url,psfrag,pstool}
\declaretheoremstyle[
spaceabove=3pt, spacebelow=3pt,
headfont=\normalfont\bfseries,
notefont=\mdseries, notebraces={(}{)},
bodyfont=\normalfont,
headformat=(\NAME\NUMBER),
postheadspace=.5em,
headpunct={}
]{mystyle}

\declaretheorem[style=mystyle,name=A,Refname=A,refname=A]{assumption}

\usepackage{mleftright}\mleftright
\date{}

\makeatother

\usepackage{babel}
\providecommand{\lemmaname}{Lemma}
\providecommand{\theoremname}{Theorem}
\providecommand{\remarkname}{Remark}
\providecommand{\corollaryname}{Corollary}
\providecommand{\examplename}{Example}

\newcommand{\noproof}[1]{}

\newcommand{\cC}{\mathcal{C}}
\newcommand{\cD}{\mathcal{D}}
\newcommand{\cE}{\mathcal{E}}

\newcommand{\cJ}{\mathcal{J}}
\newcommand{\cK}{\mathcal{K}}
\newcommand{\cL}{\mathcal{L}}

\newcommand{\cS}{\mathcal{S}}

\newcommand{\cX}{\mathcal{X}}

\newcommand{\bbE}{\mathbb{E}}

\newcommand{\bbP}{\mathbb{P}}

\newcommand{\br}{{\boldsymbol{r}}}
\newcommand{\bs}{{\boldsymbol{s}}}

\newcommand{\bu}{{\boldsymbol{u}}}
\newcommand{\bv}{{\boldsymbol{v}}}
\newcommand{\bw}{{\boldsymbol{w}}}
\newcommand{\bx}{{\boldsymbol{x}}}
\newcommand{\by}{{\boldsymbol{y}}}
\newcommand{\bz}{{\boldsymbol{z}}}

\begin{document}
\global\long\def\mult#1#2{\mu_{#1}^{#2}}
\global\long\def\mul#1{\boldsymbol{\mu}_{#1}}
\global\long\def\xmul#1{\boldsymbol{x}_{#1}}
\global\long\def\xmult#1#2{x_{#1}^{#2}}
\global\long\def\Mmul#1{\boldsymbol{M}_{#1}}
\global\long\def\hmul#1{\boldsymbol{h}_{#1}}
\global\long\def\hmult#1#2{h_{#1}^{#2}}
\global\long\def\diff#1#2{\delta_{#1}^{#2}}
\global\long\def\dif#1{\boldsymbol{\delta}_{#1}}
\global\long\def\muln#1{X_{#1}}
\global\long\def\ymul#1{\boldsymbol{y}_{#1}}
\global\long\def\ymult#1#2{y_{#1}^{#2}}
\global\long\def\zmul#1{\boldsymbol{z}_{#1}}
\global\long\def\zmult#1#2{z_{#1}^{#2}}
\global\long\def\rhomul#1{\boldsymbol{\rho}_{#1}}
\global\long\def\rhomult#1#2{\rho_{#1}^{#2}}
\global\long\def\init#1{\bar{x}^{#1}}
\global\long\def\ev{\mathbb{E}}
\global\long\def\ab{\mathcal{A}}
\global\long\def\eqlsyms#1#2{\left\langle #1,#2\right\rangle }
\global\long\def\nullsp{\operatorname{Null}}
\global\long\def\spn{\operatorname{Span}}
\global\long\def\maxr{\hat{m}}
\global\long\def\ga{\mathsf{A}}
\global\long\def\gb{\mathsf{B}}
\global\long\def\gc{\mathsf{C}}
\global\long\def\gd{\mathsf{D}}
\global\long\def\gE{\mathsf{E}}
\global\long\def\gf{\mathsf{F}}
\global\long\def\gg{\mathsf{G}}
\global\long\def\gh{\mathsf{H}}
\global\long\def\gi{\mathsf{I}}
\global\long\def\gj{\mathsf{J}}
\global\long\def\gk{\mathsf{K}}
\global\long\def\gl{\mathsf{L}}
\global\long\def\gm{\mathsf{M}}
\global\long\def\gn{\mathsf{N}}
\global\long\def\go{\mathsf{O}}
\global\long\def\gp{\mathsf{P}}
\global\long\def\gq{\mathsf{Q}}
\global\long\def\gr{\mathsf{R}}
\global\long\def\gs{\mathsf{S}}
\global\long\def\gt{\mathsf{T}}
\global\long\def\gu{\mathsf{U}}
\global\long\def\gv{\mathsf{V}}
\global\long\def\gw{\mathsf{W}}
\global\long\def\gx{\mathsf{X}}
\global\long\def\ga{\mathsf{y}}
\global\long\def\ga{\mathsf{z}}

\def\forsome{{\rm ~for~some~ }}
\let\oldforall\forall
\def\forall{{\rm ~for~all~}}
\def\forany{{\rm ~for~any~}}
\def\th{{\rm th~}}
\def\RLL{{\rm RLL}}
\def\col{{\rm col}}
\def\Var{{\rm Var}}
\def\Ham{{\rm Ham}}
\def\head{{\rm head}}
\def\tail{{\rm tail}}
\def\mid{{\rm mid}}

\newcommand{\abs}[1]{\left| #1 \right|}
\newcommand{\parenv}[1]{\left( #1 \right)}
\newcommand{\brckt}[1]{\left[#1\right]}
\newcommand{\mathset}[1]{\left\{ #1 \right\}}
\newcommand{\ccap}{\mathsf{cap}}
\newcommand{\floor}[1]{\left\lfloor #1\right\rfloor}
\newcommand{\ceil}[1]{\left\lceil #1\right\rceil}

\title{Data Deduplication with Random Substitutions}
 
\author{%
\IEEEauthorblockN{Hao Lou, Farzad Farnoud (Hassanzadeh)}
\IEEEauthorblockA{Electrical and Computer Engineering, 
University of Virginia, VA, USA. 
Email: \{haolou,farzad\}@virginia.edu}
\thanks{This paper was presented in part at ISIT 2020~\cite{lou2020data}.}
\thanks{Hao Lou is with the Department
    of Electrical and Computer Engineering, 
    University of Virginia, Charlottesville, VA, 22903, USA, (email: haolou@virginia.edu).}
\thanks{Farzad Farnoud (Hassanzadeh) is with the Department of Electrical and Computer Engineering and the Department of Computer Science, University of Virginia, Charlottesville, VA, 22903, USA, (email: farzad@virginia.edu).}
\thanks{This work was supported by NSF grant under grant no. CIF-1755773.}
}
\newcommand{\entropy}[1]{#1}
\newcommand{\longversion}[1]{#1}
\newcommand{\shortversion}[1]{}

\maketitle
\begin{abstract}
Data deduplication saves storage space by identifying and removing repeats in the data stream. Compared with traditional compression methods, data deduplication schemes are more computationally efficient and are thus widely used in large scale storage systems. In this paper, we provide an information-theoretic analysis of the performance of deduplication algorithms on data streams in which repeats are not exact. We introduce a source model in which probabilistic substitutions are considered. More precisely, each symbol in a repeated string is substituted with a given edit probability. Deduplication algorithms in both the fixed-length scheme and the variable-length scheme are studied. The fixed-length deduplication algorithm is shown to be unsuitable for the proposed source model as it does not take into account the edit probability. Two modifications are proposed and shown to have performances within a constant factor of optimal for a specific class of source models with the knowledge of model parameters. We also study the conventional variable-length deduplication algorithm and show that as source entropy becomes smaller, the size of the compressed string vanishes relative to the length of the uncompressed string, leading to high compression ratios.
\end{abstract}


\section{Introduction}
The task of reducing data storage costs is gaining increasing attention due to the explosive growth of the amount of digital data, especially redundant data~\cite{gantz2012digital,meyer2012study,el2012primary}. Data deduplication is a data reduction approach that eliminates duplicate data at the file or subfile level. Compared with traditional data compression approaches, data deduplication is more efficient when dealing with large-scale data. It has been widely used in mass data storage systems, e.g., LBFS (low-bandwidth network file system)~\cite{muthitacharoen2001low} and Venti~\cite{quinlan2002venti}. In this paper, we aim to study the performance of data deduplication algorithms from an information-theoretic point of view when repeated data segments are not necessarily exact copies.

A typical data deduplication system uses a chunking scheme to parse the data stream into multiple data `chunks'. Chunks are entered into the dictionary at the first occurrences, and duplicates are replaced by pointers to the dictionary. The chunks can be of equal length (fixed-length chunking) or of lengths that are content-defined (variable-length chunking)~\cite{manber1994finding}. The fixed-length scheme has low complexity but suffers from the boundary-shift problem: if insertions or deletions occur in a part of the data stream, then all subsequent chunks are changed because the boundaries are shifted. In the variable-length scheme, chunk breakpoints are determined using pre-defined patterns and therefore edits will not affect subsequent chunks and repeated data segments can still be identified. 

An information-theoretic analysis of deduplication algorithms was first performed by Niesen~\cite{niesen2019information}. Niesen's work introduced a source model, formalized deduplication algorithms in both fixed-length and variable-length schemes, including (conventional) fixed-length deduplication (FLD) and variable-length deduplication (VLD), and analyzed their performance. We adopt a similar strategy in this paper. The source model introduced by Niesen produces data strings that are composed of blocks, with each block being an exact copy of one of the source symbols, where the source symbols are pre-selected strings. It is often the case, however, that the copies of a block of data that is repeated many times are approximate, rather than exact. This may occur, for example, due to edits to the data, or in the case of genomic data\footnote{Repeats are common in genomic data. For example, a majority of the human genome consists of interspersed and tandem repeated sequences~\cite{lander2001}.}, due to mutations. Thus, in our source model, we add probabilistic substitutions to each block, resulting in data streams composed of approximate copies of the source symbols. 

We then analyze data deduplication algorithms over source models with probabilistic edits. For the fixed-length scheme, three algorithms: a generalization of FLD~\cite{niesen2019information} named \emph{modified fixed-length} deduplication (mFLD), a variant of mFLD named \emph{adaptive fixed-length} deduplication (AFLD), and the \emph{edit-distance} deduplication (EDD), are presented and analyzed. Due to the boundary-shift problem, algorithms in the fixed-length scheme are studied over the source model where all source symbols have the same length. We show that for mFLD, if the chunk length is not properly chosen, the average length of the compressed strings is greater than source entropy by an arbitrarily large multiplicative factor for small enough edit probability. Meanwhile, AFLD and EDD take source model parameters into account and are shown to have performances within a constant factor of optimal. For the variable-length scheme, we consider the general scenario where source symbols are of random lengths. We show that VLD can achieve large compression ratios relative to the length of the uncompressed strings.

A large number of works have studied data deduplication; see~\cite{xia2016comprehensive} for a comprehensive survey. However, the problem is not well-studied from an information-theoretic point of view. This is important because information-theoretic analysis would enable comparing the performance of deduplication algorithms with theoretical limits, under appropriate probabilistic models, and guide the development of more efficient, possibly optimal, algorithms. In addition to the seminal work by Niesen~\cite{niesen2019information}, the work~\cite{vestergaard2019generalized} also analyzed deduplication from an information-theoretical point of view but with a source model that is incompatible with the current analysis. The problem of deduplication under edit errors was also considered in~\cite{conde2018data}. While~\cite{conde2018data} focuses on performing deduplication on two files, one being an edited version of the other by insertions and deletions, we consider a single data stream with substitution errors. 

The rest of the paper is organized as follows. Notation and preliminaries are given in the next section. In Section~\ref{sec:sourcemodel}, we introduce the information source model and bound its entropy. In Section~\ref{sec:alg}, we formally state the deduplication algorithms. In Section~\ref{sec:rslts}, we summarize the main results of this paper. Bounds on the performances of algorithms in the fixed- and the variable-length schemes are derived in Section~\ref{sec:fx} and~\ref{sec:vl}, respectively. We close the paper with concluding remarks and open problems in Section~\ref{sec:con}.

\section{Preliminaries}
We consider the binary alphabet $\{0,1\}$, denoted $\Sigma$. The set of all finite strings over $\Sigma$ (including the unique empty string) is denoted $\Sigma^*$. A $j$-(sub)string is a (sub)string of length $j$. For a non-negative integer $m$, let $\Sigma^{m}$ be the set of all strings of length $m$ over $\Sigma$. For strings $\bu,\bv\in\Sigma^*$, the concatenation of $\bu$ and $\bv$ is denoted $\bu\bv$, and the concatenation of $i$ copies of $\bu$ is denoted $\bu^i$. We denote the substring of length $\ell$ starting from the $j$-th symbol of $\bu$ by $\bu_{j,\ell}$, which is also referred to as the $j$-th $\ell$-substring of $\bu$. The length of $\bu$ is denoted $\abs{\bu}$. The cardinality of a set $S$ is also denoted $\abs{S}$. For a set $T$ of strings, $\bu$ is said to be a substring of $T$ if $\bu$ is a substring of one or more strings in $T$.


In this paper, all logarithms are to the base 2. For $0\le p\le 1$, $H(p)$ denotes the binary entropy function: $p\log\parenv{\frac{1}{p}}+(1-p)\log\parenv{ \frac{1}{1-p}}$. For $0\le p,q\le 1$, $H(p,q)$ denotes the cross entropy function: $p\log \parenv{\frac{1}{q}} +(1-p)\log\parenv{\frac{1}{1-q}}$. For an event $\cE$, we use $\bar{\cE}$ to denote its complement and use $I_\cE$ to denote the indicator variable for $\cE$, which takes value 1 when $\cE$ is true, and 0 otherwise. 

The following inequalities are used frequently: for $x\in(0,1)$ and a positive integer $n$, 
\begin{align}
    \frac{1}{2}\min\left(1,nx\right)\le 1-(1-x)^n\le\min(1,nx).\label{eq:1-xton}
\end{align}

A binary string is $k$-runlength-limited ($k$-RLL)~\cite{marcus2001introduction} if it does not contain $k$ consecutive zeros, i.e., all runs of zeros in the string are of lengths less than $k$. We denote the set of binary $k$-RLL strings by $R_k$ and denote the set of binary $k$-RLL strings of length $n$ by $R_k^n$. The following lemma provides bounds on the size of $R_k^n$.
\begin{restatable}{lem}{lemRLL}\label{lem:RLL}
Let $k$ be a positive integer. The number of binary $k$-RLL strings of length $n$, $\abs{R_k^n}$, satisfies 
\begin{equation*}
    (2-\frac{1}{2^{k-2}})^n \le \abs{R_{k}^n} \le 2(2-\frac{1}{2^{k}})^n. 
\end{equation*}
\end{restatable}
Lemma~\ref{lem:RLL} is proved by induction in Appendix~\ref{app:numRLL}. By Lemma~\ref{lem:RLL}, we bound the number of binary $k$-RLL strings of lengths at most $2^k$ in the following corollary.
\begin{cor}\label{cor:RLLless2^M}
The number of binary $k$-RLL strings of lengths at most $2^k$ satisfies 
\begin{equation*}
    \sum_{n=0}^{2^k}\abs{R_k^n}\ge\sum_{n=0}^{2^k} \parenv{2-\frac{1}{2^{k-2}}}^n \ge 2^{2^k-2}. 
\end{equation*}
\end{cor}


\section{Source model}\label{sec:sourcemodel}
The source model studied in this paper extends the one described in~\cite{niesen2019information} by allowing probabilistic substitutions. The output data stream $\bs$ is a concatenation of approximate copies of source symbols. The $A$ source symbols, denoted $\gx_1,\gx_2,\ldots,\gx_A$, are iid binary strings generated in the following way. Fix a length distribution $\bbP_{l}$ over positive integers with mean $L$. For each $1 \le a \le A$, we draw $L_a$ from $\bbP_l$ and draw $\gx_a$ uniformly from $\Sigma^{L_a}$. It is important to note that, as a result of sampling with replacement, the source symbols are distributed uniformly and independently. The probability that $(\gx_1,\dotsc,\gx_A)=(\bx_1,\dotsc,\bx_A)$ given the lengths $L_a$ is $\prod_{a=1}^A\frac{1}{2^{L_a}}$
, for any set of strings $(\bx_a)$ where $\bx_a$ has length $L_a$.
So the same sequence can be drawn multiple times as source symbols. The draws are treated as separate symbols, but with the same content. We use $\cX$ to denote the source symbol alphabet, i.e., $\cX=\{\gx_1,\gx_2,\ldots,\gx_A\}$. The alphabet is thus a multiset. To simplify some of the derivations, we adopt the same assumption as~\cite{niesen2019information} that $\bbP_l$ is concentrated around its mean, specifically, $\bbP_{l}(\frac{L}{2}\le l\le 2L)=1$. 

After generating the source symbols $\gx_1,\gx_2,\ldots,\gx_A$, we generate an iid sequence of length $B$, denoted $Y_1,\ldots,Y_B$, where each $Y_b$ is an approximate copy of a randomly chosen source symbol. Specifically, for each $1 \le b \le B$, we first pick $J_b$ uniformly at random from $\{1,2,\ldots,A\}$. Next, we generate $Y_b$ by flipping each bit of $\gx_{J_b}$ independently with probability ${\delta}$, as a way of simulating edits and other changes to the data in a simple manner. The bit flipping process is referred to as a $\delta$-edit. As an example, if $\gx_{J_b}=000000$, then a possible outcome of the $\delta$-edit could be $001001$, which has probability $\delta^2(1-\delta)^4$. The data stream $\bs$ will be a concatenation of $Y_1,Y_2,\ldots, Y_B$, i.e., $\bs=Y_1Y_2\cdots Y_B$. The approximate copies $Y_1,Y_2,\ldots, Y_B$ are referred to as source blocks. The real number ${\delta}$ is referred to as the \emph{edit probability}. The entropy of this source is denoted $H(\bs)$. Note that given $\bs$, the boundaries between source blocks are not known to us. 

In this paper, we study the asymptotic regime in which $B,A,L\to\infty$ while the edit probability $\delta$ remains a constant less than $\frac{1}{2}$. We consider the situation where $A,L$ are functions of $B$ with $A\le B^{1-k_2}$ for some $0<k_2<1$ and $L=\Theta\parenv{B^{k_1}}$ for some $k_1>0$. We allow $A$ to grow large because it is reasonable to assume that as the dataset gets larger, the number of unique blocks is also higher. This necessitates $L$ to also grow large. The assumption $A\le B^{1-k_2}$ ensures that, on average, every source symbol has repeats. The polynomial relationship between $L$ and $B$ ensures that $B$ is much smaller than $2^{\Theta(L)}$. So only a small fraction of all possible strings of length $\Theta(L)$ can appear as source symbols, or edited versions of the source symbols, in the datastream. This is compatible with our intuition that only a small number of all possible strings are valid data, e.g., an image, or a piece of text or code. Furthermore, the polynomial relationship between $B$ and $L$ appears to agree with results from experiments in~\cite{el2012primary} (also referred to in~\cite{niesen2019information}) suggesting that the reasonable range for $L$ is from a few KB to a few MB ($\approx$ $10^4$ to $10^7$ bits) and for $B$ is on the order of $10^5$ to $10^9$. Nevertheless, other asymptotic regimes may also be appropriate but are left to future work for simplicity.

The following lemma provides asymptotic bounds on $H\parenv{\bs}$. Under our assumptions, $H\parenv{\bs}$ is shown to be dominated by the term $H\parenv{\delta}BL$, i.e., the main component of the entropy is the uncertainty arising from the random substitutions.
\begin{lem}\label{lem:entropy}
As $B\rightarrow\infty$, the entropy of the source model with edit probability $\delta$ satisfies
\begin{equation*}
    H(\delta)BL\le H(\bs)\le H(\delta)BL+B\log A+ A(2L+1).
\end{equation*}
\end{lem}
\begin{IEEEproof}
For the lower bound,
\begin{equation*}
    H(\bs)\ge H(\bs|\gx_{J_1},\ldots,\gx_{J_B})=\sum_{b=1}^B H\parenv{Y_b|\gx_{J_b}} = H\parenv{\delta}BL.
\end{equation*}
For the upper bound,
\begin{align*}
    H(\bs)\le& H\parenv{\bs|\gx_{J_1},\ldots,\gx_{J_B}} + H(\gx_{J_1},\ldots,\gx_{J_B}|\cX) + H\parenv{\cX}\\
    \le & H(\delta)BL+B\log A+ A(2L+1),
\end{align*}
where $H\parenv{\cX}\le A(2L+1)$ follows from the fact that for each $\gx_a$, there are at most $2^{2L+1}$ different possibilities since we assume $L_a\le 2L$.
\end{IEEEproof}

A deduplication algorithm is said to \emph{(asymptotically) achieve a constant factor of optimal} if there exists a constant $c$ (independent of $\delta$) such that $\bbE[\cL(\bs)]\le c H(\bs)$, for all $0<\delta<\frac12$ and all sufficiently large $B$, where $\cL(\bs)$ is the length of the encoding produced by the algorithm. Given our assumptions on $A,B,L$, and the result from Lemma~\ref{lem:entropy}, the entropy $H(\bs)$ is dominated by the term $H(\delta)\bbE[|\bs|]$. If $\delta$ is close to $\frac{1}{2}$, $H(\bs)$ is close to the length of the uncompressed sequence ($\bs$ is close to an iid Bernoulli(1/2) process), while if $\delta$ is close to 0, there is large gap between the two. Hence, to determine whether an algorithm achieves a constant factor of optimal, the case of small $\delta$ is especially important, which is also the case where compression is more beneficial. 

We also define the compression ratio $R=\frac{\bbE[|\bs|]}{\bbE[\cL(\bs)]}$. 
Note that if there exists a constant $c_1$ independent of $\delta$ such that $R\le c_1$, then the algorithm uses more bits than the entropy by an arbitrarily large multiplicative factor as $\delta$ goes to 0. While if $R\rightarrow \infty$ as $\delta\rightarrow 0$, then the algorithm can achieve arbitrarily large compression ratios as entropy decreases. Finally, if there exists a constant $c_2$ such that $R\ge \frac{c_2}{H(\delta)}$ for all valid $\delta$, then the algorithm achieves a constant factor of optimal.

We discuss some strategies that we use in the rest of the paper for computing $\bbE[\cL(\bs)]$. We say $\gx_{J_b}$ is the ancestor of $Y_b$ and $Y_b$ is a descendant of $\gx_{J_b}$. For each $a$, we use $Y(a)$ to denote the set $\{1 \le b \le B: J_b=a\}$ and use $Y_{\nicefrac12}(a)$ to denote the set $\{1\le b\le\ceil{B/2}: J_b=a\}$. In other words, $Y(a)$ is the set of source block indexes of the descendants of $\gx_a$ and $Y_{\nicefrac12}(a)$ is the set of source block indexes of the descendants of $\gx_a$ among the first half of source blocks.  

Note that $\bbE[|Y(a)|]=B/A$ and $\bbE[|Y_{1/2}(a)|]=B/(2A)$. We use $\cE_u$ to denote the event that $\abs{Y(a)} \le \frac{3B}{2A}$ for all $1 \le a \le A$, and use $\cE_l$ to denote the event that $\abs{Y_{\nicefrac12}(a)}\ge\frac{B}{4A}$ for all $1 \le a \le A$.
Since
\(\abs{Y(a)}=\sum_{b = 1}^B I_{J_b=a},\) 
where all summands are iid with expected value $\frac{1}{A}$, by the Chernoff bound~\cite{mitzenmacher2017probability} and the union bound,
\begin{align}
    \Pr\parenv{\cE_u} \ge 1-Ae^{-\frac{B}{10A}},\quad 
    \Pr\parenv{\cE_l} \ge 1-Ae^{-\frac{B}{16A}}.\label{eq:probeuel}
\end{align}
Given our assumption that $A\le B^{1-k_2}$, asymptotically $\frac{B}{16A}-\log A$ goes to infinity. So the probability of $\cE_u$ goes to 1 (also true for $\cE_l$). In the performance analysis of deduplication algorithms, we generally only need to consider the case in which $\cE_l$ or $\cE_u$ holds. Specifically, we use the following inequalities as bounds on $\bbE\brckt{\cL(\bs)}$:
\begin{align*}
    \bbE[\cL(\bs)] &\le \bbE[\cL(\bs)|\cE_u] + \bbE[\cL(\bs)|\bar{\cE_{u}}] \cdot\Pr(\bar{\cE_u}),\\
    \bbE[\cL(\bs)] &\ge \bbE[\cL(\bs)|\cE_l]\cdot\Pr(\cE_l) = \bbE[\cL(\bs)|\cE_l]\cdot\parenv{1-\Pr(\bar\cE_l)}.
\end{align*}
To find $\bbE[\cL(\bs)]$, we generally compute the terms $\bbE[\cL(\bs)|\cE_u]$, $\bbE[\cL(\bs)|\cE_l]$ and show that the terms $\bbE[\cL(\bs)|\bar{\cE_{u}}] \cdot\Pr(\bar{\cE_u})$ and $\bbE[\cL(\bs)|\cE_l]\cdot\Pr(\bar\cE_l)$ are asymptotically negligible, using trivial bounds on $\cL\parenv{\bs}$.

\section{Deduplication schemes}\label{sec:alg}
In this section, we formally state the deduplication algorithms, which can be regarded as mathematical abstractions of real-world deduplication systems. All algorithms are dictionary-based and composed of two stages: chunking and encoding. In particular, the conventional fixed-length deduplication (FLD) and variable-length deduplication (VLD) algorithms were formalized in~\cite{niesen2019information} and are restated here.

In FLD, the chunk length $\ell$ is fixed. Source string $\bs$ is parsed into segments
of length $\ell$, i.e.,~$\bs = \bz_1\bz_2 \cdots \bz_{C+1}$, where~$\abs{\bz_1} = \abs{\bz_2} = \cdots =\abs{\bz_C} = \ell$,~$C = \floor{\abs{\bs}/\ell}$. The substrings $\{\bz_c\}_{c=1}^{C+1}$ are collected as deduplication chunks. The encoding process starts with encoding the length of $\bs$ by a prefix-free code for positive integers, such as the Elias gamma code~\cite{elias1975universal}, to ensure that the whole scheme is prefix-free. The chunks are then encoded sequentially. Starting with $c=1$, if chunk $\bz_c$ appears for the first time, i.e., $\bz_c\ne \bz_i$ for all $i<c$, then it is encoded as the bit 1 followed by $\bz_c$ itself and is entered into the dictionary. Otherwise, when there already exists an entry in the dictionary storing the same string as $\bz_c$, it will be encoded as the bit 0 followed by a pointer to that entry. The pointer is an index of the dictionary entries and thus can be encoded by at most $\log\abs{T^{c-1}}+1$ bits, where $T^{c-1}$ denotes the dictionary just after $\bz_{c-1}$ is processed. The number of bits FLD takes to encode $\bs$ is denoted $\cL_{F}(\bs)$. It was shown in~\cite{niesen2019information} that FLD is ineffective when source symbols have different lengths. So in this paper, we study FLD (as well as its variations, mFLD and AFLD, described below) only for sources in which all source symbols have the same length. We note that such sources are not realistic except for some scenarios such as deduplication in virtual machine disk images~\cite{jin2009effectiveness}. However, the analysis of FLD and its variants is helpful for the study of VLD, described next, as it reveals important insights about the effect of chunk lengths on the performance.

\begin{exmp}\label{exmp:fld}
For $\bs = 01101101$ and $\ell=2$, the chunks generated by fixed-length chunking are $\bz_1=01,\bz_2=10,\bz_3=11,\bz_4=01$. The encoding of length $\abs{\bs}=8$ by Elias gamma coding is $0001000$. Chunks $\bz_1$, $\bz_2$ and $\bz_3$ are new chunks and thus are encoded as $101,110,111$, respectively. Chunk $\bz_4$ is a duplicate of $\bz_1$. When $\bz_4$ is processed, the dictionary contains three strings $01$, $10$ and $11$. So $\bz_3$ is encoded as $000$, where the first 0 indicates that the chunk is repeated and the following $00$ represents the first entry of the dictionary. Concatenating all components, the final encoding of $\bs$ is $0001000101110111000$. Note that after encoding terminates, the dictionary is the ordered set $\{01,10,11\}$, which appear in the encoded string as the set of chunks preceded by indicator bits with value 1. 
\end{exmp}

In VLD, a string of length $M$ (we assume $0^M$) is chosen as the marker string. The source string $\bs$ is parsed into chunks that end with the marker string. Specifically, the source string $\bs$ is parsed as $\bs=\bz_1\cdots \bz_C$, where each $\bz_c$ (except for perhaps the last one) contains a single appearance of $0^M$ at the end. We again use $\bz_1,\ldots,\bz_C$ to represent the chunks. After splitting $\bs$ into the chunks $\{\bz_c\}_{c=1}^C$, the same dictionary encoding process as in FLD is conducted. The number of bits variable-length deduplication takes to encode $\bs$ is denoted $\cL_{VL}(\bs)$.
\begin{exmp}
Consider the same string $\bs=01101101$ as Example~\ref{exmp:fld}. VLD, with marker length $M=1$, parses $\bs$ as chunks $0,110,110,1$. The length of $\bs$ is still encoded by $0001000$. Chunks $0,110,1$ are new and are encoded with $10,1110,11$, respectively. The second occurrence of $110$ is encoded by a $0$ followed by the pointer $1$. The final encoding of $\bs$ is thus $00010001011100111$.
\end{exmp}

The \emph{modified fixed-length} deduplication (mFLD) has the same encoding process as FLD but with a two-stage chunking process. In mFLD, first, the source string $\bs$ is parsed into segments of length $D$, and then, each segment is parsed into chunks of length $\ell$, where $\ell\le D$. Specifically, the source string $\bs$ is parsed as
\begin{align*}
    \bs=\bx_1\bx_2\cdots \bx_{K+1}, \quad \abs{\bx_1} = \abs{\bx_2} = \cdots = \abs{\bx_K} =D,
\end{align*} 
where $K = \floor{\abs{\bs}/D}$ and 
\begin{align*}
    \bx_k=\bz_k^1 \bz_k^2\cdots \bz_k^{N+1},\quad \abs{\bz_k^1} = \abs{\bz_k^2} = \cdots = \abs{\bz_k^N }=\ell,
\end{align*} 
with $ 1\le k \le K$, $N=\floor{D/\ell}$ ($\bx_{K+1}$ is parsed in the same way). The number of bits mFLD takes to encode $\bs$ is denoted $\cL_{mF}(\bs)$.

Note that mFLD is a generalization of FLD since with $D=\ell$, mFLD is equivalent to FLD with the same chunk length $\ell$. For FLD to perform well, the source symbols must all have the same length $L$ and the chunk length $\ell$ must also be chosen equal to $L$ to maintain synchronization between the chunks and symbols. The generalization to mFLD allows us to maintain synchronization by setting $D=L$ and frees us to choose values other than the symbol length for the chunk length $\ell$. This flexibility enables us to study the effect of chunk length, which as we will see, will provide important intuitions for more practical algorithms such as VLD. We will focus on analyzing the performance of mFLD and report that of FLD as a corollary.

The \emph{adaptive fixed-length} deduplication (AFLD) is a specialization of mFLD with source model parameters taken into account. Given $A,B,L,\delta$, AFLD is the version of mFLD with chunk length specified as $\ell = \ceil{\frac{\log(B/A)}{H(\gamma,\delta)}}$ ($\ell=D$ if $D<\ceil{\frac{\log(B/A)}{H(\gamma,\delta)}}$) for some $\gamma\in (\delta,1/2)$. AFLD thus contains two parameters $D$ and $\gamma$. Note that in practice, source model parameters can be estimated from data. We will show later that AFLD is an optimized version of mFLD. The distinction in names is made to emphasize the optimality and also for the convenience of referring to this version of the algorithm. The number of bits AFLD takes to encode $\bs$ is denoted $\cL_{AF}(\bs)$. 

\emph{Edit-distance} deduplication (EDD) extends FLD by encoding chunks relative to previously observed \emph{similar} chunks, if any. 
EDD takes the source model parameters into account and is only defined for source models with edit probability $\delta<1/4$. EDD has two parameters, chunk length $\ell$ and mismatch ratio $\beta$, where $\delta< \beta\le 1/4$. The chunking scheme is the same as in FLD, i.e., parsing the source string $\bs$ into chunks of length $\ell$, denoted $\bz_1,\bz_2,\ldots,\bz_{C+1}$. The encoding starts with a prefix-free code representing the length of the source string. Next, each chunk $\bz_c$ is encoded as the bit 1 followed by itself if no chunk has appeared before whose Hamming distance from $\bz_c$ is at most $2\beta\ell$. Otherwise, let $c'$ be the smallest index such that the Hamming distance between $\bz_{c'}$ and $\bz_c$ is $\le 2\beta \ell$. Chunk $\bz_c$ will be encoded as the bit 0 followed by a pointer to the dictionary entry where $\bz_{c'}$ is stored, along with the bits describing the mismatches between $\bz_c$ and $\bz_{c'}$. The mismatches are the indexes of positions in which $\bz_{c'}$ and $\bz_c$ differ. Since we restrict the number of mismatches to be no more than $2\beta \ell$, the mismatches can be encoded in at most~$\log \parenv{\sum_{i=0}^{\floor{2\beta \ell }} {\ell\choose i}} + 1\le H(2\beta)\ell+1$ bits. The number of bits EDD uses to store $\bs$ is denoted by $\cL_{ED}(\bs)$.

Encoding differences between similar chunks is usually used as a post-deduplication process, which spends extra computation to eliminate redundancy among distinct but similar chunks~\cite{shilane2012wan,shilane2012delta,xia2014combining,xia2014ddelta}. In this paper, we study EDD as a simple abstraction of this type of algorithms and only consider the fixed-length chunking scheme. An edit-distance based variable-length algorithm may potentially lead to better performance and be more practically important. We leave it to future consideration due to the technical challenges in the analysis, primarily arising from the facts that chunk boundaries may shift because of edits and that deriving the statistics of the number of detected copies within a certain distance does not appear readily tractable.

\section{Results}\label{sec:rslts}
In this section, we summarize  the main results of the paper. Detailed analysis and proofs of these results will be provided in the following corresponding sections.

\subsection{Modified fixed-length deduplication and its variants}
We first present results  for mFLD and its variants AFLD and FLD. Fixed-length deduplication has been shown in~\cite{niesen2019information} to not perform well when source symbols have variable lengths. So for algorithms in the fixed-length scheme, we assume $\bbP_l$ is degenerate and let the first-stage parsing length be equal to the source symbol length.

The mFLD algorithm allows us to set the chunk length $\ell$. The effect of this length is investigated in Theorems~\ref{thm:fxlbdsml},~\ref{thm:fxlbdlgl}, and~\ref{thm:fx}. For simplicity of presentation, we give detailed analysis about the theorems in Section~\ref{sec:fx} and provide corollaries here as summaries.



\begin{cor}\label{cor:fxbad}
Consider the source model in which source symbols all have length $L$. For mFLD with $D=L$, if the chunk length $\ell = o(\log B)\cup \omega(\log B)$, the compression ratio $\frac{\bbE\brckt{\abs{\bs}}}{\bbE\brckt{\cL_{mF}(\bs)}}$ is upper bounded by a universal constant for any edit probability $\delta>0$.
\end{cor}
Corollary~\ref{cor:fxbad} follows directly from Theorems~\ref{thm:fxlbdsml} and~\ref{thm:fxlbdlgl}. It characterizes the performance of mFLD when the chunk length $\ell$ is chosen too small or too large. With the chunk length improperly chosen, the average length of the compressed strings is always at least a constant factor of the original length, regardless of the edit probability $\delta$. This is not desirable for small $\delta$ since, as $\delta$ goes to 0, the entropy gets smaller and the ratio $\frac{\bbE\brckt{\cL_{mF}\parenv{\bs}}}{H\parenv{\bs}}$ grows unboundedly. It can be seen later from the proofs of Theorems~\ref{thm:fxlbdsml} and~\ref{thm:fxlbdlgl} that when the chunk length is chosen too small, the dictionary becomes so large that the pointers become of similar lengths to the chunks. On the other hand, when the chunk length is chosen too large, repeats can not be identified and deduplication thus fails. It is therefore important to pick a suitable chunk length when implementing deduplication algorithms in practice.

If we pick $\ell=L$, mFLD becomes FLD with chunk length equal to source symbol length, which was shown in~\cite{niesen2019information} to be asymptotically optimal on sources with fixed symbol length and no edits. However, in the case when edit probability $\delta$ is nonzero, since we assume $L=\Theta\parenv{B^{k_1}}$, Corollary~\ref{cor:fxbad} implies that the compression ratio of FLD is bounded and the gap between FLD and entropy can be arbitrarily large, as stated in the next corollary.
\begin{cor}\label{cor:FLD}
Consider the source model in which source symbols all have length $L$. For FLD, with chunk length $L$, the compression ratio $\frac{\bbE\brckt{\abs{\bs}}}{\bbE\brckt{\cL_{F}(\bs)}}$ is upper bounded by a universal constant for any edit probability $\delta>0$.
\end{cor}

AFLD has its chunk length chosen adapted to source parameters and is shown in Theorem~\ref{thm:fx} to be nearly optimal. The following corollary is a summary of Theorem~\ref{thm:fx}.

\begin{cor}\label{cor:AF}
For any edit probability $\delta\in(0,\frac{1}{2})$ and any $a>1$, there exists $\delta<\gamma<\frac{1}{2}$ such that
\begin{equation*}
    \frac{\bbE[\cL_{AF}(\bs)]}{H(\bs)}\le \frac{a(1+k_1)}{k_2}\parenv{1+o(1)}.
\end{equation*}
\end{cor}
With $k_1,k_2$ being fixed constants, the preceding corollary states that AFLD achieves a constant factor of optimal for any edit probability $\delta$. Thus, to achieve high compression ratio, deduplication algorithm parameters, especially the chunk length, should be chosen based on the data. In practice, it can thus be beneficial to first obtain an estimate of the parameters of the data and then apply deduplication with algorithm parameters properly chosen. A fixed chunk length is unlikely to be universally effective for all datasets.

\subsection{Edit-distance deduplication}
The edit-distance deduplication is studied in Theorem~\ref{thm:edDist} and shown to achieve performance a constant factor of optimal.
\begin{restatable*}{thm}{edDist}\label{thm:edDist}
Consider the source model in which source symbols have the same length $L$ and the edit probability is $\delta<\frac{1}{4}$. The performance of edit-distance deduplication with chunk length $\ell= L$ and mismatch ratio $\beta$ satisfies 
\begin{align*}
    1\le \frac{\bbE[\cL_{ED}(\bs)]}{H(\bs)}\le \frac{H(2\beta)}{H(\delta)}\parenv{1+o(1)},  \quad \text{as } B\rightarrow \infty,
\end{align*}
for any $\delta<\beta\le \frac{1}{4}$.
\end{restatable*}

Note that for any $\delta<\frac{1}{4}$, we can always find $\beta$ larger than but close enough to $\delta$ such that $\frac{H(2 \beta)}{H(\delta)}$ is upper bounded by a constant value. With such choices of $\beta$, the preceding theorem states that $\bbE\brckt{\cL_{ED}(\bs)}$ is at most a constant factor of $H(\bs)$. As an example, let $\beta = \min\parenv{\frac{3\delta}{2},\frac{1}{4}}$. The ratio $\frac{H(2\beta)}{H(\delta)}$ is upper bounded by 
\begin{align*}
    \frac{H(2\beta)}{H(\delta)} \le \frac{H(\min\parenv{3\delta,1/2})}{H(\delta)} \le 3,
\end{align*}
where the last inequality follows from the fact that $\frac{H(3p)}{H(p)}\le 3$ for all $p\le \frac{1}{3}$ and $H(\frac{1}{6})\le \frac{1}{2}$. Hence, EDD also achieves a constant factor of optimal, as formalized in the following corollary.

\begin{cor}\label{cor:edDistCor}
Consider the source model in which source symbols have the same length $L$ and edit probability $\delta<\frac{1}{4}$. There exists a mismatch ratio $\beta$ such that the performance of EDD with chunk length $\ell=L$ satisfies
\begin{equation*}
    \frac{\bbE[\cL_{ED}(\bs)]}{H(\bs)}\le \frac{H(3\delta)}{H(\delta)}(1+o(1)) \le 3\parenv{1+o(1)}.
\end{equation*}
\end{cor}
We note however that EDD is more complex than AFLD as it identifies chunks that are within a certain Hamming distance. 

\subsection{Variable-length deduplication}
Similar to the algorithms in the fixed-length scheme, the performance of VLD depends on the chunk length. In VLD, the chunk length is controlled by the length $M$ of the marker (the expected chunk length is approximately $2^M$). The effect of $M$ on the performance is studied in Theorems~\ref{thm:vllbdsmallM},~\ref{thm:vllbdlargeM},~\ref{thm:vllbdhugeM} and~\ref{thm:vlubd}, in Section~\ref{sec:vl}.




As a summary of Theorems~\ref{thm:vllbdsmallM},~\ref{thm:vllbdlargeM}, and~\ref{thm:vllbdhugeM}, we first present the following corollary, showing that an inappropriate choice of $M$ leads to poor performance.
\begin{cor}\label{cor:vld}
Consider the source model with edit probability $\delta$ and variable-length deduplication with marker length $M$. If $2^M=o(\log B)\cup \omega(\log B)$, the compression ratio $\frac{\bbE\brckt{\abs{\bs}}}{\bbE\brckt{\cL_{VL}(\bs)}}$ is upper bounded by a universal constant for any edit probability $\delta>0$.
\end{cor}

\begin{figure}[h]
    \centering
    \includegraphics[width=0.45\textwidth]{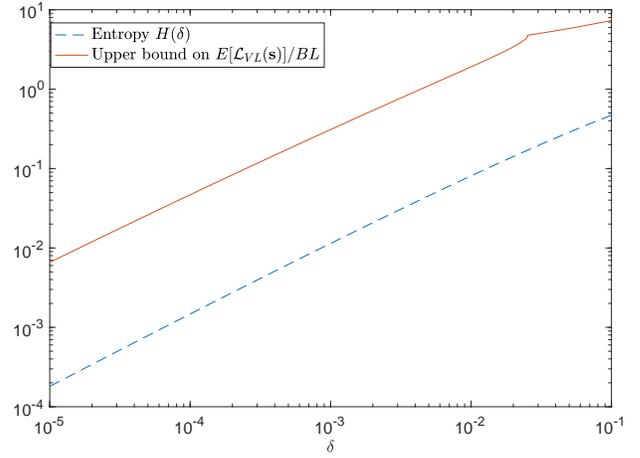}
    \caption{Upper bound on $\frac{\bbE[\cL_{VL}(\bs)]}{BL}$ and $H(\delta)$ vs the edit probability $\delta$ with $A=L=B^{\nicefrac12}$, as $\delta$ ranges from $10^{-5}$ to $10^{-1}$.}
    \label{fig:vlup}
\end{figure}

We also show that a well-chosen marker length $M$ can lead to arbitrarily large compression ratios as edit probability $\delta$ approaches 0.
\begin{restatable*}{thm}{vlubd}\label{thm:vlubd}
Consider the source model with edit probability $\delta<\frac12$. For any $\gamma\in(\delta,1/2)$, the performance of variable-length deduplication with marker length $M$ such that $2^M=\Theta\parenv{\log\parenv{B/A}}$ satisfies
\begin{align}
     \bbE&\brckt{\cL_{VL}\parenv{\bs}} \le\nonumber\\
     & \parenv{\! 12e^{-c_M}\parenv{c_M+1} \!+\! 4H\parenv{\gamma,\delta}\frac{(1+k_1)}{k_2}c_M\!}BL\parenv{1+o(1)}, \label{eq:vlubdcm}
\end{align}    
as $B\rightarrow\infty$, where $c_M=\frac{\log\parenv{B/A}}{H\parenv{\gamma,\delta}2^{M+1}}$.
\end{restatable*}
We perform the following analysis for minimizing the upper bound given by Theorem~\ref{thm:vlubd}. For any given $c>0$, there exists an integer value for $M$ such that $c\le c_M\le 2c$. For this $M$, \eqref{eq:vlubdcm} is upper bounded by
\begin{align*}
    \parenv{12e^{-c}\parenv{c+1} + 8H\parenv{\gamma,\delta}\frac{(1+k_1)}{k_2}c}BL\parenv{1+o(1)},
\end{align*}
since $e^{-c}(c+1)$ is decreasing in $c$ when $c>0$. We can always find $\gamma$ such that $H\parenv{\gamma,\delta}\le 2H\parenv{\delta}$. Such $\gamma$ gives
\begin{align}
    &\quad\bbE\brckt{\cL_{VL}\parenv{\bs}} \nonumber\\
    &\le \parenv{12e^{-c}\parenv{c+1} + 16H\parenv{\delta}\frac{(1+k_1)}{k_2}c}BL\parenv{1+o(1)}.\label{eq:vlubddelta}
\end{align}
Let $h=4H\parenv{\delta}\frac{(1+k_1)}{3k_2}$. Upper bounding the above expression is equivalent to upper bounding the function $f(c) = e^{-c}(c+1) + hc$, $c\in\parenv{0,+\infty}$. If $h< e^{-1}$, then $f(c)$ has a local minimum at $c=-W_{-1}\parenv{-h}$, where $W_{-1}$ is the lower branch of the Lambert $W$ function. If $h\ge e^{-1}$, then $f(c)$ is monotonically increasing in $\parenv{0,+\infty}$. Therefore, $c=-W_{-1}(-\min\parenv{e^{-1},h})$ provides an upper bound on $f(c)$. As an example, for $A=L=B^{\nicefrac12}$ (i.e., $k_1=k_2=\frac{1}{2}$), Figure~\ref{fig:vlup} shows the upper bound given by \eqref{eq:vlubddelta} with $c=-W_{-1}(-\min\parenv{e^{-1},h})$, as well as $H(\delta)$, as $\delta$ ranges from $10^{-5}$ to $10^{-1}$.

Note that $h\le e^{-1}$ holds for small enough $\delta$. When this holds, the upper bound \eqref{eq:vlubddelta} can be rewritten as 
\begin{align*}
    &\quad \bbE\brckt{\cL_{VL}\parenv{\bs}} \\
    &\le \parenv{12e^{-c}\parenv{c+1} + 16H\parenv{\delta}\frac{(1+k_1)}{k_2}c}BL\parenv{1+o(1)}\\ 
    &\le 12e^{-c}\parenv{c^2+c+1}BL\parenv{1+o(1)},
\end{align*}
where $c=-W_{-1}\parenv{-4H\parenv{\delta}\parenv{1+k_1}/\parenv{3 k_2}}$. Hence the upper bound on the normalized expected compressed length approaches 0 as ${\delta}$ approaches 0. This means that 
as the entropy becomes smaller, the compression ratio grows if the length of the marker is chosen appropriately. 
In particular, it can be seen that the proper length of the marker depends on ${\delta}$, which represents the degree of variability between the copies. 

Large compression ratios when entropy is small is desirable and variable-length deduplication achieves this. However, it can be shown and also observed in Figure~\ref{fig:vlup} that the upper bound of the ratio $\bbE[\cL_{VL}(\bs)]/H(\bs)$ given by Theorem~\ref{thm:vlubd} increases as $\delta$ decreases. Therefore, despite the large compression ratios, the gap to entropy may become large for small $\delta$. Determining whether this is indeed the case or the bound provided here is loose is left to future work.

\section{Deduplication in the Fixed-length Scheme}\label{sec:fx}
In this section, we study the performances of the deduplication algorithms in the fixed-length scheme. It is pointed out by~\cite{niesen2019information} that when all source symbols have the same length and there are no edits, FLD with knowledge of the symbol length can parse data strings in a way that chunk boundaries align with source block boundaries (by setting the chunk length equal to source block length) and achieve asymptotically optimal performance under mild conditions. However, when symbols have different lengths, the loss of synchronization leads to poor performance. For instance,~\cite{niesen2019information} considered the scenario in which there are $A=2$ source symbols, with the source symbol length distribution $\bbP_l$ assigning equal probability to $L$ and $L+1$ (here $L$ is an independent parameter rather than the expected value of $\bbP_l$) and with $B=3L$ source blocks. FLD with chunk length $\ell=L$ was shown to satisfy $\frac{\bbE\brckt{\cL_{F}(\bs)}}{H(\bs)} \ge \Omega(B)$. In the case where copies are not exact, the question of interest is then whether fixed-length deduplication can still perform well when chunk boundaries align with repeat boundaries. To answer this question, we need to ensure that the two groups of boundaries are aligned. So we consider only source models where source symbols all have the same length $L$ ($\bbP_l$ is degenerate). 

We first study in detail the performance of mFLD and then specialize the results to FLD. The first-stage parsing length of mFLD (including AFLD) and the chunk length of EDD are both assumed to be equal to $L$. 


We present a lemma that will be used frequently. For positive integers $m,\ell$ and $\delta\in\parenv{0,\frac{1}{2}}$, define 
\begin{equation*}
    \cS_{\delta}(\ell,m)=\sum_{t=0}^{\ell}{\ell\choose t}\min\parenv{1,m\delta^t(1-\delta)^{\ell-t}}.
\end{equation*}

\begin{restatable}{lem}{wapponecol}\label{lem:wapp1col}
Let $\br$ be a string drawn uniformly at random from $\Sigma^{\ell}$. Let $\br_1,\br_2,\ldots,\br_m$ be $m$ iid descendants of $\br$ by $\delta$-edit and let~$\br_{[m]}=\{\br_1,\br_2,\ldots,\br_m\}$. For any $\bw\in \Sigma^{\ell}$, let $\bw\in \br_{[m]}$ denote the event that $\bw=\br_i$ for some $i$. Then
\begin{align}
    \frac{1}{2}\frac{\cS_{\delta}(\ell,m)}{2^\ell}
    \le\Pr\parenv{\bw \in \br_{[m]}}\le
    \frac{\cS_{\delta}(\ell,m)}{2^\ell},\label{eq:Prwinrm}
\end{align}
and thus the expected number of unique strings in $\br_{[m]}$ is bounded between $\frac{1}{2}\cS_{\delta}\parenv{\ell,m}$ and $\cS_{\delta}\parenv{\ell,m}$.

Furthermore, $\cS_{\delta}\parenv{\ell,m}$ takes the following values for different values of $\ell$ and $m$:
\begin{itemize}
    \item If $\ell\ge \frac{\log m}{H(\delta)}$, then 
    \begin{align}
        \cS_{\delta}\parenv{\ell,m} \ge\frac{1}{4}m. \label{eq:sellmlargeell}
    \end{align}
In particular if $\ell \ge \frac{\log m}{\log(\frac{1}{1-\delta})}$, then 
\begin{align}
    \cS_{\delta}\parenv{\ell,m} =m.\label{eq:sellmhugeell}
\end{align}

\item If $\ell\le \frac{\log m}{H(\frac{1}{2},\delta)}$, then
\begin{align}
    \cS_{\delta}\parenv{\ell,m} \ge 2^{\ell-1}. \label{eq:sellmsmallell}
\end{align}
In particular if $\ell\le \frac{\log m}{ \log(\frac{1}{\delta})}$, then
\begin{align*}
    \cS_{\delta}\parenv{\ell,m} =  2^\ell. 
\end{align*}

\item For any $\delta<\delta'<\frac{1}{2}$,
\begin{align}
    \cS_{\delta}\parenv{\ell,m}\le 2^{\ell H(\delta')}+ m2^{-\ell D(\delta'||\delta)}. \label{eq:cSdelta'}
\end{align}
In particular if $\ell = \frac{\log m}{H(\delta',\delta)}$, then 
\begin{align}
       \cS_{\delta}\parenv{\ell,m}\le2^{\ell H(\delta')}+ m2^{-\ell D(\delta'||\delta)}=2^{\ell H(\delta')+1}. \label{eq:sellmmiddleell}
\end{align}
\item For any values of $\ell$ and $m$,
\begin{align}
    \cS_{\delta}\parenv{\ell,m} \le \min\parenv{2^\ell,m}. \label{eq:sellmubd}
\end{align}
\end{itemize}
\end{restatable}
The proof of Lemma~\ref{lem:wapp1col} is presented in Appendix~\ref{app:wapp1col}.

\subsection{Modified and adaptive fixed-length deduplication}\label{subsec:mfld}
We show that, even with knowledge of the source symbol length, if the chunk length is not properly chosen, mFLD encodes $\bs$ with a constant number of bits per symbol regardless of $\delta$. Therefore, the ratio $\frac{\bbE\brckt{\cL_{mF}(\bs)}}{H(\bs)}$ can be arbitrarily large for small $\delta$. Meanwhile for AFLD, with the adaptive chunk length $\ell = \ceil{\frac{\log(B/A)}{H(\gamma,\delta)}}$, the ratio $\frac{\bbE\brckt{\cL_{AF}(\bs)}}{H(\bs)}$ is shown to be upper bounded by a constant for all $\delta$ and for $\gamma$ properly chosen.

\begin{figure}[h]
    \centering
    \scalebox{0.9}{\begin{tikzpicture}[
		node distance = 0cm,
		gn/.style={minimum width=.3cm, minimum height=0.5cm, shape=rectangle,draw,font=\small,inner sep=0mm},]
	\node (y1name) [draw=none]{$Y_1$};
	\node (y1) [gn,right = of y1name, xshift = 0.5cm, minimum width= 8cm]{};
	\node (vdot1)[draw = none, below = of y1, yshift = -0cm]{$\vdots$};
	\node (ybname)[draw= none, below = of y1name, yshift = -0.9cm]{$Y_b$};
	\node (yb)[gn, right = of ybname, xshift = 0.5cm, minimum width = 8 cm]{};
	\node (yBname)[draw = none, below = of ybname, yshift=-0.9cm]{$Y_B$};
	\node (yB)[gn, right = of yBname, xshift = 0.5cm, minimum width = 8cm]{};
	
	\node (vdot2)[draw = none, below = of yb, yshift = -0cm]{$\vdots$};
    \node (z11) [gn, minimum width = 2cm, right = of y1name,xshift = 0.5cm]{$Z_1^1$};
	\node (z21) [gn, minimum width = 2cm, right = of z11]{$Z_2^1$};
	\node (cdot1)[gn, minimum width = 1cm, right = of z21]{$\cdots$};
	\node (zC1)[gn, minimum width = 2cm, right = of cdot1]{$Z_C^1$};
	\node (zcp11) [gn, draw = none,right = of zC1]{$Z_{C+1}^1$};
	
	\node (z1b) [gn, minimum width = 2cm, right = of ybname,xshift = 0.5cm]{$Z_1^b$};
	\node (z2b) [gn, minimum width = 2cm, right = of z1b]{$Z_2^b$};
	\node (cdot2)[gn, minimum width = 1cm, right = of z2b]{$\cdots$};
	\node (zCb)[gn, minimum width = 2cm, right = of cdot2]{$Z_C^b$};
	\node (zcp1b) [gn, draw=none,right = of zCb]{$Z_{C+1}^b$};
	
	\node (z1B) [gn, minimum width = 2cm, right = of yBname,xshift = 0.5cm]{$Z_1^B$};
	\node (z2B) [gn, minimum width = 2cm, right = of z1B]{$Z_2^B$};
	\node (cdot3)[gn, minimum width = 1cm, right = of z2B]{$\cdots$};
	\node (zCB)[gn, minimum width = 2cm, right = of cdot3]{$Z_C^B$};
	\node (zcp1B) [gn, draw=none,right = of zCB]{$Z_{C+1}^B$};
	
	  \draw[decorate,decoration={brace,amplitude=5pt,raise=1pt},yshift=-1pt] ([xshift = -1cm]z11.north) -- ([xshift = 1cm]z11.north) node [black,midway, yshift = 0.5cm] 
    {\footnotesize $\ell$};
	\draw[decorate,decoration={brace,amplitude=5pt,raise=1pt},yshift=-1pt] ([xshift = 1cm]zC1.north) -- ([xshift = 4cm]y1.north) node [black,midway, yshift = 0.5cm] 
    {\footnotesize $r$};
	
	\end{tikzpicture}}
    \caption{Modified fixed-length chunking with segment length $D=L$ and chunk length $\ell$.}
    \label{fig:dbfxck}
\end{figure}
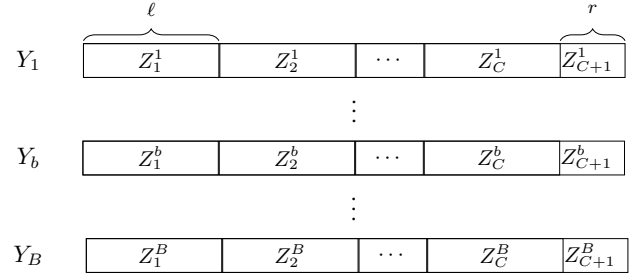

Consider the two-stage parsing of $\bs$ with $D=L$. The length-$D$ segments after the first-stage parsing are exactly the source blocks $Y_1,Y_2,\ldots,Y_B$. Let $C = \floor{L/\ell}$ and $r = L-C\ell$. Each $Y_b$, $1\le b\le B$, is then parsed into chunks $Z_1^b,Z_2^b,\ldots,Z_{C+1}^b$ with $\abs{Z_c^b}=\ell$ for all $c\le C$ and $\abs{Z_{C+1}^b}=r$ (see Figure~\ref{fig:dbfxck}). If we also divide each source symbol $\gx_a$ into substrings of length $\ell$ as $\gx_a= U_1^a U_2^a\cdots U_{C+1}^a$, then for all $1\le c\le C+1$, $\{Z_c^b\}_{b\in Y(a)}$ are iid $\delta$-edit descendants of $U_c^a$. 

Before performing a detailed evaluation of the algorithm, let us first provide a rough analysis for a special case, which will provide some insights into the general problem. Suppose the alphabet $\cX$ only has a single symbol $\gx$ of length $L$, whose $\ell$-prefix is denoted by $U_1$. We consider encoding \emph{only} the set $Z_1^1,Z_1^2,\dotsc,Z_1^B$, where each $Z_1^b$ is a descendant of $U_1$  by $\delta$-edit. The expected size of the dictionary, i.e., the number of distinct $\ell$-strings in $\{Z_1^1,Z_1^2,\dotsc,Z_1^B\}$, by Lemma~\ref{lem:wapp1col} is approximately
\begin{equation}\label{eq:dict_size_toy}
S:=\cS_{\delta}(\ell,B)=\sum_{t=0}^\ell \min\left(\binom{\ell}{t},\binom{\ell}{t}B\delta^t(1-\delta)^{\ell-t}\right).
\end{equation}
We can interpret~\eqref{eq:dict_size_toy} as follows. At a given distance $t$ from $U_1$, there are $\binom{\ell}{t}$ sequences of length $\ell$. Further, if we generate $B$ sequences, the expected number of sequences at distance $t$ is $\binom{\ell}{t}B\delta^t(1-\delta)^{\ell-t}$. The number of sequences in the dictionary at distance $t$ is then approximated by the minimum of the two terms. (This analysis of $S$ is helpful whenever $\cS_\delta(\cdot,\cdot)$ appears in the sequel as well.)

We would like $S$ to be small enough that $\log S\ll \ell$ (so that pointers to the dictionary have much smaller lengths than the sequences being encoded) and $S\ll B$ (so that each sequence in the dictionary is repeated many times).\footnote{Note that the size of the dictionary, and hence the length of the pointers, vary as the encoding progresses; we ignore this fact for now and approximate pointer lengths based on the final size of the dictionary.} As $t$ ranges from $0$ to $\ell$ in the sum in~\eqref{eq:dict_size_toy}, the term $\binom{\ell}{t}$ attain its maximum at $t\simeq\ell/2$ while the second term inside the $\min$ attains its maximum at $t\simeq\ell\delta$. We investigate which term determines the behavior of the sum. Let $\ell=\frac{\log B}{H(\gamma,\delta)}$ for a constant $0\le\gamma\le 1$. Note that since $\delta<\frac{1}{2}$, $H(\gamma,\delta)$ and $\ell$ are increasing and decreasing functions of $\gamma$, respectively. With this choice, $B\delta^t(1-\delta)^{\ell-t}\ge 1$ for $t\le\ell\gamma$ and $B\delta^t(1-\delta)^{\ell-t}\le 1$ for $t\ge\ell\gamma$.
\begin{itemize}
    \item If $\gamma<\delta$, then $B\delta^{\delta\ell}\parenv{1-\delta}^{(1-\delta)\ell}<1$, and  $S\ge\sum_{t=\ceil{\gamma\ell}}^{\ell} \binom{\ell}{t}B\delta^t(1-\delta)^{\ell-t}\ge B(1-2^{-\ell D(\gamma||\delta)})$. In this case, almost all $Z_1^b$ are distinct and thus not compressible.
    \item If $\gamma=\delta$, then $\ell=\frac{\log B}{H(\delta)}$, and  $S\ge\frac{B}{4}$ by \eqref{eq:sellmlargeell}. In this case, a constant fraction of $Z_1^b$ are distinct and thus not compressible.
    \item If $\gamma\ge 1/2$, then $\ell\le\frac{\log B}{H(\frac{1}{2},\delta)}$, and $S\ge 2^{\ell-1}$ by \eqref{eq:sellmsmallell}. In this case, due to the fact that $\ell$ is chosen too small, the dictionary is so large that pointers to the dictionary are as long as the chunks and there is no compression gain.
    \item If $\delta<\gamma<1/2$,  then by \eqref{eq:sellmmiddleell},
    \begin{equation*}
        S\le 2^{\ell H(\gamma)+1}.
    \end{equation*} 
    Hence, pointers have an approximate length of $\ell H(\gamma)$ and are smaller than $\ell$ by a factor of $\frac{1}{H(\gamma)}$. Furthermore, each sequence is repeated approximately $2^{\ell D(\gamma||\delta)}$ times since  $B=2^{\ell H(\gamma,\delta)}$. The number of bits required to encode the dictionary is $2\ell 2^{\ell H(\gamma)}$, which is negligible compared to $B\ell$, the length of the uncoded sequences since $\gamma\neq \delta$. Hence, we can encode $\{Z_1^1,\dotsc,Z_1^B\}$ using essentially $B\ell H(\gamma)$ bits, achieving a compression ratio of $\frac{1}{H(\gamma)}$.
\end{itemize}
This analysis highlights that $\ell$ should be chosen appropriately to avoid a large dictionary or a situation in which there are no repetitions in the sequence. If these conditions are satisfied, then we can successfully deduplicate the data, as shown rigorously in Theorem~\ref{thm:fx} for AFLD. 

Now we return to the general setting. It can be seen from the description of mFLD that the compressed string is composed of two parts: the bits used to encode the chunks at their first occurrences and the bits used to encode repeated chunks by pointers to the dictionary. For both parts, our first step is to compute the expected size of the dictionary, i.e., the number of distinct chunks, for which we present Lemma~\ref{lem:wappearingrid} and Lemma~\ref{lem:dicsizefx}.



\begin{restatable}{lem}{lemwappearingrid}\label{lem:wappearingrid}
Suppose $K$ strings of length $n$ are chosen independently and uniformly from $\Sigma^{n}$. Assume each string produces at least $m_1$ and at most $m_2$ descendants by $\delta$-edits. For any string $\bw$ with $\abs{\bw}= n$, let $G_{\bw}$ denote the event that $\bw$ equals one or more descendants. Then
\begin{multline*}
    \frac{1}{2}\min\parenv{1,\frac{1}{2} K\frac{\cS_{\delta}\parenv{n,m_1}}{2^{n}}} \\
    \le \Pr\parenv{G_\bw} \le \\\min\parenv{1,K\frac{\cS_\delta\parenv{n,m_2}}{2^{n}}}.
\end{multline*}
\end{restatable}
The proof of Lemma~\ref{lem:wappearingrid} is presented in Appendix~\ref{app:wappearingrid}. This lemma considers the probability of observing a string $\bw$ when multiple random strings produce $\delta$-edit descendants simultaneously. This setting models exactly our source string generation process where the $A$ source symbols correspond to $K$ random strings, and the source blocks correspond to the $\delta$-edit descendants. In particular, $\cE_u$ being true corresponds to $m_2=\frac{3B}{2A}$ and $\cE_l$ being true corresponds to $m_1=\frac{B}{4A}$.


Let $T_{F}^1(\bs)$ denote the dictionary after all chunks of $\bs$ are processed, i.e., $T_{F}^1(\bs)$ contains all distinct strings in $\{Z_c^b\}_{b,c}$. Let $T_{F}^{\nicefrac12}(\bs)$ denote the dictionary immediately after all chunks in the first half of $\bs$, i.e., $Y_1Y_2\cdots Y_{\ceil{B/2}}$, are processed. We apply Lemma~\ref{lem:wappearingrid} to find bounds on the sizes of $T_F^1(\bs)$ and $T_F^{\nicefrac{1}{2}}(\bs)$ in the following lemma.
\begin{restatable}{lem}{corwappearingrid}\label{lem:dicsizefx}
Consider the two-stage fixed-length chunking process with first-stage parsing length $D=L$ and chunk length $\ell$. The dictionary sizes $T_F^1(\bs)$ and $T_F^{\nicefrac{1}{2}}(\bs)$ satisfy
\begin{align}
&\bbE\brckt{\abs{T_{F}^1(\bs)} |\cE_u}
  \le \min\parenv{2^\ell,AC\cS_\delta\left(\ell,\frac{3B}{2A}\right)} + B,\nonumber\\
  &\bbE\brckt{\abs{T_{F}^{\nicefrac12}(\bs)} |\cE_l} \ge \frac{1}{2}\min\parenv{2^\ell,\frac{1}{2}AC\cS_\delta\left(\ell,\frac{B}{4A}\right)}. \label{eq:TF1/2lbd}
\end{align}
\end{restatable}
The proof of Lemma~\ref{lem:dicsizefx} is presented in Appendix~\ref{app:dicsizefx}.


Next, we show using Lemma~\ref{lem:dicsizefx} that if $\ell$ is chosen too small relative to the scale of the system, then mFLD spends a constant number of bits per symbol.
The proof strategy is as follows: with $\ell$ small enough, the term $\min\parenv{2^\ell,\frac{1}{2}AC\cS_{\delta}\parenv{\ell,\frac{B}{4A}}}$ in \eqref{eq:TF1/2lbd} equals 1, which makes $\bbE\brckt{\abs{T_F^{\nicefrac{1}{2}}(\bs)}|\cE_l}$ greater than $2^{\ell-1}$. Therefore, when encoding duplicated chunks using pointers, each pointer takes approximately $\ell$ bits and there is no compression gain.
\begin{restatable}{thm}{fxlbdsml}\label{thm:fxlbdsml}
Consider the source model in which source symbols have the same length $L$. For mFLD with first-stage parsing length $D=L$ and chunk length $\ell$, if $\ell2^\ell = O(AL)$ or \(\ell\le\frac{\log(B/A)-2}{H(\frac{1}{2},\delta)}\), then
\begin{align*}
    \bbE[\cL_{mF}(\bs)] \ge \frac{1}{12}BL(1+o(1)), \quad \text{as } B\rightarrow \infty,
\end{align*}
where the $o(1)$ term is independent of $\delta$.
\end{restatable}
\begin{IEEEproof}
We first claim, to be proved later, that if $\ell2^\ell = O(AL)$ or \(\ell\le\frac{\log(B/A)-2}{H(\frac{1}{2},\delta)}\), then 
\begin{align}
    \bbE\brckt{\abs{T_{F}^{\nicefrac12}(\bs)}|\cE_l} \ge 2^{\ell-1}.\label{eq:E1/2ge2^ell-1}
\end{align}
It follows from Markov's inequality that
\begin{equation*}
    \Pr\parenv{2^\ell-\abs{T_{F}^{\nicefrac12}(\bs)} \ge \frac{3}{4}\cdot2^\ell|\cE_l}\le \frac{\frac{1}{2}\cdot 2^\ell}{\frac{3}{4}\cdot 2^\ell} = \frac{2}{3},
\end{equation*}
which is equivalent to
\begin{equation}
\Pr\parenv{\abs{T_{F}^{\nicefrac12}(\bs)}\ge \frac{2^\ell}{4}|\cE_l} \ge \frac{1}{3}. \label{eq:problbddicsize}    
\end{equation}
Next, we consider the second half of $\bs$, $Y_{\ceil{B/2}+1}\cdots Y_B$. There are $\floor{B/2}C$ chunks of length $\ell$, and encoding each of them takes at least either $\ell$ or $\log\abs{T_{F}^{\nicefrac12}(\bs)}$ bits plus an additional bit indicating whether the chunk is stored in full or represented by a pointer. So in total, we need at least
\begin{align*}
    \parenv{\min\parenv{\ell,\log\abs{T_{F}^{\nicefrac12}(\bs)}} +1} \cdot \floor{\frac{B}{2}}C
\end{align*}
bits. It follows that for $B$ sufficiently large, 
\begin{align*}
    \bbE\brckt{\cL_{mF}(\bs)|\cE_l} &\ge \bbE\brckt{\parenv{\min\parenv{\ell,\log\abs{T_{F}^{\nicefrac12}(\bs)}} +1} \cdot \floor{\frac{B}{2}}C|\cE_l} \\
    & \ge \frac{1}{3}\parenv{\min\parenv{\ell,\log\frac{2^\ell}{4}}+1}\cdot\floor{\frac{B}{2}}C \\
    &\ge \frac{BL}{12}\parenv{1+o(1)},
\end{align*}
where the second inequality follows from \eqref{eq:problbddicsize}. 

Finally, since \eqref{eq:probeuel} gives that $\Pr\parenv{\cE_l} = 1+o(1)$, we get
\begin{align*}
       \bbE\brckt{\cL_{mF}(\bs)} \ge \bbE\brckt{\cL_{mF}(\bs)|\cE_l}\Pr\parenv{\cE_l}\ge \frac{BL}{12}\parenv{1+o(1)}.
\end{align*}

It remains to prove the claim:  $\bbE\brckt{\abs{ T_{F}^{\nicefrac12}(\bs)}|\cE_l}\ge 2^{\ell-1}$ when $\ell2^\ell = O(AL)$ or \(\ell\le\frac{\log(B/A)-2}{H(\frac{1}{2},\delta)}\).
Consider the case when $\ell 2^\ell = O(AL)$. For sufficiently large $B$ (and thus $A$ and $L$), $\frac{B}{4A}\ge \frac{4\ell2^\ell}{AL}$. Therefore, by Lemma~\ref{lem:dicsizefx}, 
\begin{align*}
    \bbE\brckt{\abs{T_{F}^{\nicefrac12}(\bs)}|\cE_l} &\ge\frac{1}{2}\min\parenv{2^\ell,\frac{1}{2}AC\cS_\delta\left(\ell, \frac{B}{4A}\right)} \\
    &\ge \frac{1}{2}\min\parenv{2^\ell,\frac{1}{2}AC\cS_\delta\left(\ell, \frac{4\ell2^\ell}{AL}\right)},
\end{align*}
where the last inequality follows from the fact that $\cS_\delta\parenv{\ell,m}$ is non-decreasing in $m$. By \eqref{eq:sellmhugeell}, if $m(1-\delta)^\ell\le 1$, then $\cS_\delta(\ell,m)=m$. Since asymptotically $\frac{4\ell2^\ell}{AL}(1-\delta)^\ell\le 1$,
\begin{align*}
    \bbE\brckt{ \abs{T_{F}^{\nicefrac12}(\bs)}|\cE_l} \ge 2^{\ell-1}\min\parenv{1,\frac{4\ell C}{2L}}\ge 2^{\ell-1},
\end{align*}
where the last step follows from the fact that $C\ge \frac{L}{2\ell}$.

When \(\ell\le\frac{\log(B/A)-2}{H(\frac{1}{2},\delta)}\), again by Lemma~\ref{lem:dicsizefx},
\begin{align*}
    \bbE\brckt{\abs{T_{F}^{\nicefrac12}(\bs)}|\cE_l} &\ge  \frac{1}{2}\min\parenv{2^\ell,\frac{1}{2}AC\cS_\delta\left(\ell,\frac{B}{4A}\right)}\\&\ge 2^{\ell-1}\min\parenv{1,\frac{AC}{4} }\ge2^{\ell-1},
\end{align*}
where the second inequality follows from \eqref{eq:sellmsmallell} that when $\ell\le\frac{\log m}{H(\frac{1}{2},\delta)}$, $\cS_{\delta}\parenv{\ell,m}\ge 2^{\ell-1}$.
\end{IEEEproof}

The preceding theorem shows that when $\ell$ is chosen too small, the size of the dictionary will be of order $2^\ell$. Specifically, if $\ell2^\ell=O(AL)$, the number of distinct $\ell$-substrings in the source alphabet is already of order $2^\ell$. If $\ell\le\frac{\log(B/A)-2}{H(\frac{1}{2},\delta)}$, then the $\delta$-edits are able to produce almost all $\ell$-strings instead of only producing strings that are on the $\delta\ell$ Hamming sphere.

In the next theorem, we show that if $\ell$ is chosen too large, then mFLD again spends a constant number of bits per symbol. The proof strategy is to show that if $\ell$ is chosen too large, then almost every chunk is distinct, thus making the source string incompressible.

\begin{restatable}{thm}{fxlbdlgl}\label{thm:fxlbdlgl}
Consider the source model in which source symbols have the same length $L$. For mFLD with first-stage parsing length $D=L$ and chunk $\ell$, if \(\ell\ge\frac{\log(B/A)-2}{H(\delta)}\), then
\begin{align*}
\bbE\brckt{\cL_{mF}(\bs)}\ge\frac{1}{128}BL(1+o(1)),\quad \text{as }B\rightarrow\infty,
\end{align*}
where the $o(1)$ term is independent of $\delta$.
\end{restatable}

\begin{IEEEproof}
When $\ell\le\frac{\log(B/A)-2}{H(\frac{1}{2},\delta)}$, 
\begin{align*}
    \bbE\brckt{\abs{T_{F}^{\nicefrac12}(\bs)} |\cE_l} &\ge \frac{1}{2}\min\parenv{2^\ell,\frac{1}{2}AC\cS_\delta\parenv{\ell,\frac{B}{4A}}} \\
    & \ge 2^{\ell-1}\min\parenv{1,\frac{1}{2}AC\cdot \frac{1}{4}\cdot\frac{B}{4A2^\ell}}\\
   & = 2^{\ell-1}\min\parenv{1,\frac{BC}{32\cdot 2^\ell}},
\end{align*}
where the first inequality follows from Lemma~\ref{lem:dicsizefx} and the second from \eqref{eq:sellmlargeell}.

In the case where $1\le \frac{BC}{32\cdot 2^\ell}$ and hence $\bbE\brckt{\abs{ T_{F}^{\nicefrac12}(\bs)}|\cE_l}\ge 2^{\ell-1}$, the proof follows from the discussion that follows \eqref{eq:E1/2ge2^ell-1}. So it remains to consider the case when $\frac{BC}{32\cdot2^\ell} \le 1$, i.e.,  \begin{align*}
    \bbE\brckt{\abs{T_{F}^{\nicefrac12}(\bs)}|\cE_l} \ge 2^{\ell-1}\cdot\frac{BC}{32\cdot 2^\ell} = \frac{BC}{64}.
\end{align*}

Since it takes $\ell+1$ bits to store distinct chunks in the dictionary, 
\begin{align*}
    \bbE\brckt{\cL_{mF}(\bs)|\cE_l} &\ge (\ell+1)\bbE\brckt{\abs{T_F^{\nicefrac{1}{2}}(\bs)}|\cE_l}
    = \ell\frac{B\floor{L/\ell}}{64}  
    \\&\ge \frac{1}{64}B\max\parenv{\ell,L-\ell} 
     \ge \frac{1}{128}BL.
\end{align*}
The desired result thus follows again from $\bbE\brckt{\cL_{mF}(\bs)} \ge\bbE\brckt{\cL_{mF}(\bs)|\cE_l} \Pr\parenv{\cE_l}$ and the fact that $\Pr\parenv{\cE_l}=1+o(1)$.
\end{IEEEproof}

In the Results section, Theorems~\ref{thm:fxlbdsml} and~\ref{thm:fxlbdlgl} imply Corollary~\ref{cor:fxbad}, which shows that choosing $\ell$ in $o(\log B)$ or $\omega(\log B)$ results in poor performance, and Corollary~\ref{cor:FLD}, which shows that FLD cannot compress the sequences effectively.



Next, we show that with the adapted chunk length, AFLD can achieve performance within a constant factor of optimal.

\begin{restatable}{thm}{fx}\label{thm:fx}
Consider the source model in which source symbols have the same length $L$. The performance of AFLD with $D = L$ and $\ell = \ceil{\frac{\log\parenv{B/A}}{H\parenv{\gamma,\delta}}}$ satisfies
\begin{align*}
     1\le \frac{\bbE[\cL_{AF}(\bs)]}{H(\bs)} \le \frac{1+k_1}{k_2}\cdot\frac{H(\gamma,\delta)}{H(\delta)}\cdot(1+o(1)), 
\end{align*}
as $B\rightarrow \infty$, for any $\gamma\in(\delta,\frac{1}{2})$. 
\end{restatable}

\begin{IEEEproof}
We first note that the length of $\bs$ can be encoded in at most $2\log(\abs{\bs})+3$ bits with Elias gamma coding.

The number of bits used to encode chunks at their first occurrences is upper bounded by $\abs{T_F^1\parenv{\bs}}(\ell+1)$ since chunks are all of lengths less than or equal to $\ell$. Consider the upper bound on $\bbE\brckt{\abs{ T_{F}^1(\bs)}|\cE_u}$ in Lemma~\ref{lem:dicsizefx}. Note that by \eqref{eq:cSdelta'} and $\frac{B}{A}\le 2^{\ell H\parenv{\gamma,\delta}}$ with our choice of $\ell$,
\begin{align*}
    \cS_{\delta}\parenv{\ell,\frac{3B}{2A}}
    &\le  2^{\ell H(\gamma)} + \frac{3B}{2A}2^{-\ell D (\gamma||\delta)} \le \frac{5}{2}\cdot2^{\ell H(\gamma)}.
\end{align*}
It follows that
\begin{align}
    \bbE&\brckt{\abs{ T_{F}^1(\bs)}|\cE_u}(\ell+1)\nonumber\\&\le \parenv{\min\parenv{2^\ell,AC\cS_\delta\parenv{\ell,\frac{3B}{2A}}} +B} \parenv{\ell+1}\nonumber\\
    &\le\min\parenv{2^\ell,\frac{5AC}{2}\cdot2^{\ell H(\gamma)}}\parenv{\ell+1} + B\parenv{\ell+1} \nonumber\\
    & \le \frac{5AL}{2\ell}\cdot2^{\ell H(\gamma)}\cdot(\ell+1) +B\parenv{\ell+1}\nonumber\\
    &= \frac{5}{2}AL\parenv{\frac{B}{A}}^{H(\gamma)/H(\gamma,\delta)}\parenv{1+\Theta\parenv{\frac{1}{\log(B/A)}}} \nonumber\\
    &\quad + \Theta\parenv{B\log B} \nonumber \\
    & = o(BL),\label{eq:fxubdlck}
\end{align}
where the last equality follows from $\frac{H\parenv{\gamma}}{H\parenv{\gamma,\delta}}<1$ and thus $B^{\frac{H\parenv{\gamma}}{H\parenv{\gamma,\delta}}}A^{1-\frac{H\parenv{\gamma}}{H\parenv{\gamma,\delta}}}=o(B)$.


Next, we derive an upper bound on the number of bits used by pointers for encoding repeated chunks. There are $(C+1)B$ chunks and the number of bits needed for encoding one pointer is at most $\log (BL)+1$. So in total, the number of bits we need is at most
\begin{align}
    (C+1)B(\log (BL)+1) &\le (L+\ell)B\frac{\log(BL)+1}{\ell} \nonumber\\
    &\le \frac{BL}{\ell}\log(BL)\parenv{1+O\parenv{\frac{1}{\log B}}} \nonumber  \\
    & \le H(\gamma,\delta) BL \cdot\frac{\log(BL)}{\log (B/A)}. \label{eq:fxubdpt}  
\end{align}

Combining \eqref{eq:fxubdlck}, \eqref{eq:fxubdpt}, and including the number of bits used for encoding the length of $\bs$ by Elias coding, we get
\begin{align*}
\bbE\brckt{\cL_{AF}(\bs)|\cE_u} &\le H(\gamma,\delta)BL\cdot\frac{\log(BL)}{\log (B/A)} + o(BL)\\
&\le H\parenv{\gamma,\delta}BL\frac{1+k_1}{k_2} \parenv{1+o(1)},
\end{align*}
by noting that $\frac{\log (BL)}{\log (B/A)}\le \frac{1+k_1}{k_2}\parenv{1+o(1)}$.

On the complement of $\cE_u$, the number of bits needed for storing the dictionary is at most $2BL$ since the lengths of chunks in total is at most $BL$ and there are at most $BL$ chunks. The number of bits for encoding repeated chunks by pointers is at most $BL(\log(BL) +1)$. It follows that 
\begin{multline*}
    \bbE\brckt{\cL_{AF}(\bs)|\bar\cE_u}\Pr\parenv{\bar\cE_u} \\
    \le \parenv{2ABL+2\log (BL)+3}\log (BL) e^{-\frac{B}{10A}} = o(1).
\end{multline*}

The desired result thus follows from
\begin{multline*}
    \bbE\brckt{\cL_{AF}(\bs)} \\
    = \bbE\brckt{\cL_{AF}(\bs)|\cE_u}\Pr\parenv{\cE_u} + \bbE\brckt{\cL_{AF}(\bs)|\bar\cE_u}\Pr\parenv{\bar\cE_u},
\end{multline*}
and the fact that $\Pr\parenv{\cE_u}=1+o(1)$.
\end{IEEEproof}
For any $\delta<\frac{1}{2}$ and $a>1$, we can find $\gamma$ in the range $(\delta,\frac{1}{2})$ such that $H(\gamma,\delta)/H(\delta)\le a$. It thus follows from Theorem~\ref{thm:fx} that adaptive fixed-length deduplication can compress the sequence within a constant factor of the entropy, as stated in Corollary~\ref{cor:AF} in the Results section.


\subsection{Edit-distance deduplication}
Next, we study the edit-distance deduplication algorithm. EDD identifies positions in which the current chunk and previously observed similar chunks differ. We show that with chunk length being equal to source symbol length, EDD can achieve a constant factor of optimal.

\edDist

\begin{IEEEproof}
With $\ell=L$, the $B$ source blocks, $Y_1,\ldots,Y_B$, are parsed as chunks. We know that each $Y_b$ is a descendant of one of the source symbols. Let $\cE_d$ denote the event that every source block $Y_b$ is within Hamming distance $\beta L$ from its ancestor. By the Chernoff bound, the probability that more than $\beta L$ symbols of a source symbol are flipped in a $\delta$-edit is at most $2^{-D\parenv{\beta||\delta} L}$. We then apply the union bound and get $\Pr(\cE_d)\ge 1-B2^{-D\parenv{\beta||\delta} L}$.

When $\cE_d$ holds, the source blocks are covered by $A$ Hamming balls of radius $\beta L$. Therefore, with mismatch ratio $\beta$, the dictionary is of size at most $A$, and takes $A(L+1)$ bits to store. The pointer length is thus upper bounded by $\log A +1$. The difference with the referenced chunk can be encoded in at most $H\parenv{2\beta }L +1$ bits. Including the $2\log (BL)+3$ bits for encoding $\abs{\bs}$ at the beginning, we get 
\begin{align*}
      \bbE[\cL_{ED}(\bs)|\cE_d]
      &\le 2\log (BL)+3+A(L+1) \\&\qquad+ (1+\log A +1+H(2\beta )L+1)B \\
      &= H(2\beta)BL + o(BL).
\end{align*}

When the complement of $\cE_d$ holds, we trivially upper bound dictionary size by $B$. It follows that
\begin{align*}
    \bbE[\cL_{ED}(\bs)|\bar{\cE_d}]&\le 2\log(BL)+3+B(L+1) \\
    &\qquad + \parenv{1+\log B + 1+H(2\beta)L +1}B \\
    &\le 2BL.
\end{align*}

Thus, 
\begin{align*}
    \bbE[\cL_{ED}(\bs)]&=\Pr(\cE_d)\bbE[\cL_{ED}(\bs)|\cE_d] + \Pr(\bar{\cE_d})\bbE[\cL_{ED}(\bs)|\bar{\cE_d}]\\
    &\le H(2\beta)BL(1+o(1)) + 2B^2L2^{-D\parenv{\beta||\delta}L}\\
     &=H(2\beta)BL\parenv{1+o(1)},
\end{align*}
where the term $2B^2L 2^{-D\parenv{\beta||\delta}L}$ is absorbed into the $o(1)$ term since $D\parenv{\beta||\delta}>0$.
\end{IEEEproof}

The theorem is used in the Results section to establish that EDD performs within a constant factor of entropy in Corollary~\ref{cor:edDistCor}. 

\section{Deduplication in the Variable-length Scheme}\label{sec:vl}
In this section, we study the variable-length deduplication algorithm, which is more widely applicable than the algorithms in the fixed-length scheme and does not require the source symbol lengths to be the same or known. In the previous section, we saw that for AFLD to achieve optimality, the chunk length should be adapted to the source. Similarly for VLD, the performance depends on chunk lengths which in turn depend on the length of the marker $M$. 

Before presenting the detailed analysis, we provide some insights on how the marker length $M$ affects the distribution of chunk contents. In variable-length chunking, the chunks (except perhaps the last one) end with the marker string $0^M$. We write $\bs = U_1 0^M U_20^M\cdots 0^M U_N$, where each $U_n, n<N$, is either empty or of the form $\bu 1$ for some $M$-RLL string $\bu$. We can approximately treat $\bs$ as a Bernoulli(1/2) process for now. The lengths of strings $U_n$ are thus equivalent to the stopping time in an infinite-length Bernoulli(1/2) process untill the beginning of the first occurrence of $0^M$, which is of expected length approximately $2^M$. The behavior of VLD with marker length $M$ is thus similar to that of mFLD with chunk length $2^M$. When $M$ is chosen so small that the number $N$ of chunks becomes much larger than the total number of strings of lengths around $ 2^M$, the dictionary becomes exhaustive and pointers have similar lengths to chunks. When $M$ is chosen too large, most of $U_1,\ldots,U_N$ are distinct and thus not compressible. In the following, we study in detail how $\bbE\brckt{\cL_{VL}(\bs)}$ varies for different values of $M$.

Similar to the fixed-length schemes, the dictionary size is an essential first-step in computing $\bbE\brckt{\cL_{VL}(\bs)}$. To determine the expected dictionary size, we again start with the probability of occurrences of chunks. However, now the chunks are of different lengths and the occurrences are not restricted to a fixed set of positions. So we bound the probability of occurrences of a chunk by the probability of occurrences of certain substrings. Specifically, we consider strings of the forms $10^M\bu10^M$ or $0^M\bu10^M$ ($\bu\in R_M$): Except the first and the last chunks, the probability of occurrence of chunk $\bu10^M$ is greater than the probability of occurrence of a substring $10^M\bu 10^M$ since the prefix $10^M$ always marks an ending of the previous chunk; similarly, the probability of occurrence of chunk $\bu10^M$ is less than or equal to the probability of occurrence of a substring $0^M\bu 10^M$ since any occurrences of chunk $\bu10^M$ must follow a $0^M$ which is the ending marker of the previous chunk. 

Let $\bw\in Y_1^{B}$ denote the event that $\bw$ appears as a substring of $Y_b$ for some $1\le b\le B$ and let $\bw\in Y_1^{B/2}$ denote the event that $\bw$ appears as a substring of $Y_b$ for some $1\le b\le \ceil{B/2}$.\footnote{Here we only consider string/chunk occurrences inside source blocks and leave the study of strings/chunks that occur across the boundaries of source blocks for later.} We first present in Lemmas~\ref{lem:wappearingridsub},~\ref{lem:wappearlbdubd} and~\ref{lem:wappearlbdbigellvl} two lower bounds on $\bw\in Y_1^{B/2}$ and an upper bound on $\bw\in Y_1^B$.

\begin{restatable}{lem}{lemwappearingridsub}\label{lem:wappearingridsub}
Suppose $K$ strings of length $n$ are chosen independently and uniformly from $\Sigma^{n}$. Assume each string produces at least $m_1$ and at most $m_2$ descendants by $\delta$-edits. For any string $\bw$ with $\abs{\bw}\le n$, let $H_{\bw}$ denote the event that $\bw$ appears as a substring of one or more descendants. Then,
\begin{multline*}
    \frac{1}{2}\min\parenv{1,\frac{1}{2}\left\lfloor{\frac{n}{\abs{\bw}}}\right\rfloor K\frac{\cS_{\delta}\parenv{\abs{\bw},m_1}}{2^{\abs{\bw}}}} \\
    \le \Pr\parenv{H_\bw} \le\\ \min\parenv{1,\parenv{n-\abs{\bw}+1}K\frac{\cS_\delta\parenv{\abs{\bw},m_2}}{2^{\abs{\bw}}}}.
\end{multline*}
\end{restatable}
The proof of Lemma~\ref{lem:wappearingridsub} is presented in Appendix~\ref{app:wappearingridsub}. Similar to Lemma~\ref{lem:wappearingrid}, the setting described in Lemma~\ref{lem:wappearingridsub} matches the model for the generation of source strings. This time, we allow string $\bw$ to be any substring of the descendants because chunks can now be in any position of the source string. Note that Lemma~\ref{lem:wappearingridsub} is also a generalization of Lemma~\ref{lem:wappearingrid}.

Next, we use Lemma~\ref{lem:wappearingridsub} to bound the probability of $\bw\in Y_1^{B/2}$ and $\bw\in Y_1^B$.
\begin{restatable}{lem}{wappearlbdubd}\label{lem:wappearlbdubd}
Consider the source model with edit probability $\delta$. For any string $\bw\in\Sigma^*$ with $\abs{\bw}\le 2L$, 
\begin{align*}
    \Pr(\bw\in Y_1^B|\cE_u)\le \min\parenv{1,2AL\frac{\cS_{\delta}\parenv{\abs{\bw},\frac{3B}{2A}}}{2^{\abs{\bw}}}}.
\end{align*}
For any string $\bw\in\Sigma^*$ with $\abs{\bw}\le \ceil{\frac{1}{2}L}$,
\begin{align}
    \Pr\parenv{\bw\in Y_1^{B/2}|\cE_l} \ge \frac{1}{2}\min\parenv{1,\frac{AL}{8\abs{\bw}}\frac{\cS_{\delta}\parenv{\abs{\bw},\frac{B}{4A}}}{2^{\abs{\bw}}}}.\label{eq:wappearlbd}
\end{align}
\end{restatable}
The proof of Lemma~\ref{lem:wappearlbdubd} is presented in Appendix~\ref{app:wappearlbdubd}. Although Lemma~\ref{lem:wappearlbdubd} holds for any string $\bw$, we will later restrict $\bw$ to be of the forms $10^M\bu10^M$ or $0^M\bu10^M$. 
 
Next, we consider another lower bound as an alternative to \eqref{eq:wappearlbd} for the cases when $\bw$ is of larger lengths. From the proofs of Lemmas~\ref{lem:wappearingridsub} and~\ref{lem:wappearlbdubd}, the lower bound \eqref{eq:wappearlbd} is obtained by only taking into account the possibilities of $\bw$ appearing in non-overlapping positions of each $Y_b$. Lemma~\ref{lem:wappearlbdbigellvl} considers every possible substring of $Y_b$ to be equal to $\bw$ and gets the lower bound by the inclusion-exclusion principle and turns out to be more accurate for $\bw$ with large lengths. Note that Lemma~\ref{lem:wappearlbdbigellvl} directly considers $\bw$ to be of the form $10^M\bu10^M$ and the bound is given in the form of a summation. 
\begin{restatable}{lem}{wappearlbdbigellvl}\label{lem:wappearlbdbigellvl}
Consider the source model with edit probability $\delta<\frac12$. For any $n$ such that $\frac{\log(B/A)-2}{H(\delta)}\le n+2M+2\le \frac{L}{4}$,  
\begin{multline*}
    \sum_{\bu\in R_M^n} \Pr\parenv{ 10^M\bu 10^M  \in Y_1^{B/2}|\cE_l}
    \ge \\\frac{BL}{2^7\cdot 2^{2M+2}}\cdot \parenv{1-\frac{1}{2^{M-1}}}^n - \frac{3B^2L^2}{2^{n+2M+2}}.
\end{multline*}
\end{restatable}
The proof of Lemma~\ref{lem:wappearlbdbigellvl} is presented in Appendix~\ref{app:wappearlbdbigellvl}. 

After characterizing the probabilities of strings (and thus chunks) occurring, we consider in Lemma~\ref{lem:numlck2hf} the number of chunks. Let $C_{VL}^M(\bs)$ denote the number of chunks of length over $2^{M-4}$ in $Y_{\ceil{B/2}+1}\cdots Y_B$ for variable-length chunking with marker length $M$. We show that when $2^M=o(L)$, with high probability, $C_{VL}^M(\bs)$ is of order $\abs{\bs}/2^M$.   

\begin{restatable}{lem}{numlcktohf}\label{lem:numlck2hf}
Consider the source string $\bs=Y_1Y_2\ldots Y_B$. When $2^M=o(L)$, for $B,L$ sufficiently large,
\begin{align*}
    \Pr\parenv{C_{VL}^M(\bs) \ge \frac{1}{4}\cdot\floor{\frac{B}{2}}\parenv{\frac{L}{2^{M+8}}-1}} \ge \frac{5}{6}.  
\end{align*}
\end{restatable}
The proof of Lemma~\ref{lem:numlck2hf} is presented in Appendix~\ref{app:numlck2hf}. It can be seen from the proof that Lemma~\ref{lem:numlck2hf} can be extended to the case when $\cE_l$ holds since each source block $Y_b$ by itself is still a Bernoulli(1/2) process. Therefore, the following corollary holds.
\begin{cor}\label{cor:numlck2hf}
When $2^M=o(L)$, for $B,L$ sufficiently large,
\begin{align*}
    \Pr\parenv{C_{VL}^M(\bs) \ge \frac{1}{4}\cdot\floor{\frac{B}{2}}\parenv{\frac{L}{2^{M+8}}-1} |\cE_l} \ge \frac{5}{6}.  
\end{align*}
\end{cor}


Next, we use Lemmas~\ref{lem:wappearlbdubd},~\ref{lem:wappearlbdbigellvl} and Corollary~\ref{cor:numlck2hf} to bound $\bbE\brckt{\cL_{VL}(\bs)}$ from below. As marker length $M$ takes different values, different lower bounds of $\bbE\brckt{\cL_{VL}(\bs)}$ are presented in Theorems~\ref{thm:vllbdsmallM},~\ref{thm:vllbdlargeM} and~\ref{thm:vllbdhugeM}. Let $T_{VL}^1(\bs)$ denote the dictionary when all chunks in $\bs$ are processed and let $T_{VL}^{\nicefrac12}(\bs)$ denote the dictionary immediately after chunks in $Y_1\cdots Y_{\ceil{B/2}}$ are processed.

We first show in Theorem~\ref{thm:vllbdsmallM} that similar to the fixed-length schemes, small values for $M$ lead to an oversized dictionary. 

\begin{restatable}{thm}{vllbdsmallM}\label{thm:vllbdsmallM}
Consider the source model with edit probability $\delta$ and the variable-length deduplication algorithm with marker length $M$. If $ 2^M = o(\log B)$, then
\begin{align*}
\bbE[\cL_{VL}(\bs)]\ge \frac{1}{3\cdot 2^{16}}BL\parenv{1 + o(1)}, \quad \text{as } B\rightarrow\infty,
\end{align*}
where the $o(1)$ term is independent of $\delta$.
\end{restatable}

\begin{IEEEproof}
We show that with high probability, $\abs{T_{VL}^{\nicefrac12}(\bs)}$ is of the order $2^{2^M}$. So encoding each chunk in $Y_{\ceil{B/2}+1}\cdots Y_B$ takes number of bits either equal to the chunk length or pointer length $2^M$. We then show using Lemma~\ref{lem:numlck2hf} that the length of the compressed string is a constant fraction of $BL$.

If a string $\bw$ of the form $\bw=10^M\bu10^M$, $\bu\in R_M$, occurs as a substring of some data block $Y_b, b\le \ceil{\frac{B}{2}}$, then $\bu10^M$ must be contained in $T_{VL}^{\nicefrac12}(\bs)$. For any $\bw=10^M\bu10^M$ with $\abs{\bu}\le 2^M$, by Lemma~\ref{lem:wappearlbdubd},
\begin{align}
    \Pr\parenv{\bw\in Y_1^{B/2}|\cE_l} &\ge \frac{1}{2}\min\parenv{1,\frac{AL}{8\abs{\bw}}\frac{\cS_{\delta}\parenv{\abs{\bw},\frac{B}{4A}}}{2^{\abs{\bw}}}} \nonumber\\
    &\ge \frac{1}{2}\min\parenv{1,\frac{AL}{8\abs{\bw}}} \ge  \frac{1}{2}, \label{eq:winY1BgEl}
\end{align}
where the second inequality follows from $\abs{\bw}\le 2^M+2M+2=o(\log B)$ and the property that $\cS_{\delta}\parenv{\ell,m}=2^\ell$ if $m\delta^\ell\ge 1$. 

Denote the set of all $M$-RLL strings of lengths less than $2^M$ by $R_M^{\le 2^M}$. Let $\zeta=\left|\left\{\bu\in R_M^{\le2^M}:10^M\bu 10^M\in Y_1^{B/2}\right\}\right|$. Then \eqref{eq:winY1BgEl} gives $\bbE[\zeta|\cE_l]\ge |R_M^{\le 2^M}|/2$ and thus $\bbE[|R_M^{\le 2^M}|-\zeta|\cE_l]\le \frac{|R_M^{\le 2^M}|}{2}$. By Markov inequality, $\Pr(|R_M^{\le 2^M}|-\zeta\ge 3|R_M^{\le 2^M}|/4)\le \frac23$ and thus $\Pr(\zeta>|R_M^{\le 2^M}|/4)\ge \frac13$. Noting that $|T_{VL}^{\nicefrac{1}{2}}|\ge \zeta$ and $|R_M^{\le 2^M}|\ge 2^{2^M-2}$ by Corollary~\ref{cor:RLLless2^M}, we get
\begin{align}
    \Pr\parenv{\abs{T_{VL}^{\nicefrac12}(\bs)} \ge 2^{2^M-4}|\cE_l}\ge\frac{1}{3}.\label{eq:vllbddic} 
\end{align}

For each chunk in $Y_{\ceil{B/2}+1}\cdots Y_B$ of length at least $2^{M-4}$, we need at least either $2^{M-4}$ or $\log\abs{T_{VL}^{\nicefrac12}(\bs)}$ bits. So by Corollary~\ref{cor:numlck2hf} and inequality \eqref{eq:vllbddic},
\begin{align*}
    \bbE\brckt{\cL_{VL}(\bs)|\cE_l} &\ge \bbE\brckt{\min\parenv{2^{M-4},\log\abs{T_{VL}^{\nicefrac12}(\bs)}}\cdot C_{VL}^M(\bs)|\cE_l} \\
    &\ge \parenv{1-\frac{2}{3}-\frac{1}{6}}\min\parenv{2^{M-4},2^M-4}\\
    &\quad \cdot \frac{1}{4}\floor{\frac{B}{2}}\parenv{\frac{L}{2^{M+8}}-1}\\
    &\ge\frac{BL}{3\cdot 2^{16}}\parenv{1+o(1)}.
\end{align*}
The desired result follows from \[\bbE\brckt{\cL_{VL}(\bs)} \ge \bbE\brckt{\cL_{VL}(\bs)|\cE_l} \Pr\parenv{\cE_l}\] and $\Pr\parenv{\cE_l}=1+o(1)$.
\end{IEEEproof}

We then show in Theorems~\ref{thm:vllbdlargeM} and~\ref{thm:vllbdhugeM} that an oversized $M$ leads to a large number of distinct chunks, each of which needs to be encoded in full and thus compression becomes ineffective. In particular, Theorem~\ref{thm:vllbdlargeM} covers the case when $2^M$ is of larger order than $\log B$ but still much smaller than the expected source symbol length $L$. Theorem~\ref{thm:vllbdhugeM} considers the case when $2^M = \Omega(L)$, and therefore a large number of chunks can be of lengths close to or even larger than the expected source symbol length.
\begin{restatable}{thm}{vllbdlargeM}\label{thm:vllbdlargeM}
Consider the source model with edit probability $\delta$ and the variable-length deduplication algorithm with marker length $M$. If $2^M=\omega(\log B)\cap o(L)$, then
\begin{align*}
\bbE[\cL_{VL}(\bs)]\ge \frac{1}{2^{10}e^{2}}BL\parenv{1 +o(1)}, \quad \text{as } B\rightarrow\infty,
\end{align*}
where the $o(1)$ term is independent of $\delta$.
\end{restatable}

\begin{IEEEproof}
We show that if $2^M$ is in $\omega(\log B) $ and $o(L)$, the sum of the lengths of distinct chunks is a constant fraction of $\abs{\bs}$.

Each new chunk is encoded as a bit 1 followed by itself. Given $\cE_l$, the expected number of bits needed for encoding distinct chunks is greater than or equal to
\begin{align}
    \bbE&\brckt{\sum_{\bv\in T_{VL}^1(\bs)}\parenv{\abs{\bv}+1}|\cE_l}\nonumber\\&=  \sum_{\bv\in \Sigma^* } \Pr\parenv{\bv\in T_{VL}^1\parenv{\bs}|\cE_l}\parenv{\abs{\bv}+1} \nonumber\\
    &\ge \sum_{\bu\in R_M}\Pr\parenv{10^M\bu 10^M\in Y_1^{B/2}|\cE_l}\parenv{\abs{\bu}+M+2}.\quad  \label{eq:dicprob}
\end{align}
As a lower bound, we consider $M$-RLL strings with lengths in the range $\brckt{2^M, \ceil{\parenv{2^ML}^{\nicefrac12}}}$. Since asymptotically we have $2^M\ge \frac{\log(B/A)-2}{H(\delta)}$, we apply Lemma~\ref{lem:wappearlbdbigellvl} on \eqref{eq:dicprob} and get
\begin{align*}
&\sum_{\ell=2^M}^{\ceil{\parenv{2^ML}^{\nicefrac12}}}\sum_{\bu\in R_M^\ell}\Pr\parenv{10^M\bu10^M\in Y_1^{B/2} |\cE_l}\\
&\cdot\parenv{\ell+M+1}\\
    \ge& \sum_{\ell=2^{M}}^{\ceil{\parenv{2^ML}^{\nicefrac12}}} \left(\frac{BL}{2^7\cdot 2^{2M+2}} \parenv{1-\frac{1}{2^{M-1}}}^\ell \right.\left.- \frac{3 B^2L^2}{2^{\ell+2M+2}}\right)\\
    &\cdot\parenv{\ell+M+1}\\
    \ge & \sum_{\ell=2^{M}}^{\ceil{\parenv{2^ML}^{\nicefrac12}}} \left(\frac{BL}{2^7 \cdot 2^{2M+2}} \parenv{1-\frac{1}{2^{M-1}}}^\ell\ell \right)- \frac{3 B^2L^4}{2^{2^M}} \\
    \ge & \frac{BL}{2^7\cdot 2^{2M+2}}2^{2(M-1)}\parenv{\frac{2^{M}-1}{2^{M-1}} +1} e^{-2}\parenv{1+o(1)} \\
    & - \frac{3B^2L^4}{2^{2^M}}\\
    \ge &\frac{BL}{2^{10}e^2} \parenv{1+o(1)} - \frac{3B^2L^4}{2^{2^M}}\\
    =&\frac{BL}{2^{10}e^2} \parenv{1+o(1)}, \quad \text{as } B\rightarrow\infty, 
\end{align*}
where the second inequality follows from $\parenv{2^ML}^{\nicefrac12}+M+1\le L$ and the equality follows from $2^M=\omega\parenv{\log B}$. The second to last inequality follows from applying summation \eqref{eq:sumlbl} in Appendix~\ref{app:sum2} with $a=2^M, b = \ceil{\parenv{2^M L}^{\nicefrac12}}, \beta = 2^{M-1}$ and noting that $ \frac{1}{2^{M-1}}\ceil{\parenv{2^M L}^{\nicefrac12}}=\omega(1)$.


Thus,
\begin{align*}
    \bbE\brckt{\cL_{VL}(\bs)|\cE_l} &\ge \bbE\brckt{\sum_{\bv\in T_{VL}^1(\bs)}\parenv{\abs{\bv}+1}|\cE_l} \\
    &\ge \frac{BL}{2^{10}e^2}\parenv{1+o(1)},
\end{align*}
and the desired result follows from \[\bbE\brckt{\cL_{VL}(\bs)} \ge \bbE\brckt{\cL_{VL}(\bs)|\cE_l} \Pr\parenv{\cE_l}\] and $\Pr\parenv{\cE_l}=1+o(1)$.
\end{IEEEproof}

Next, we present a lemma that will be used in the proof of Theorem~\ref{thm:vllbdhugeM}.
\begin{restatable}{lem}{lemidsstrins}\label{lem:idsstrins}
Consider the source string $\bs = Y_1Y_2\cdots Y_B$, with each $Y_b$ being a descendant of source symbol $\gx_{J_b}$. For any integer $h$ and any pairs of integers $(b_1,b_2),(i_1,i_2)$, the probability of $Y_{b_1}$ and $Y_{b_2}$ having identical substrings of length $h$ starting at positions $i_1$ and $i_2$, respectively, is
\begin{align*}
    \Pr\parenv{\parenv{Y_{b_1}}_{i_1,h} = \parenv{Y_{b_2}}_{i_2,h}} = \frac{1}{2^h},
\end{align*}
if $J_{b_1}\ne J_{b_2}$ or $i_1\ne i_2$.
\end{restatable}
The proof of Lemma~\ref{lem:idsstrins} is presented in Appendix~\ref{app:insstrins}.

\begin{restatable}{thm}{vllbdhugeM}\label{thm:vllbdhugeM}
Consider the source model with edit probability $\delta$ and the variable-length deduplication algorithm with marker length $M$. If $2^M=\Omega(L)$, then
\begin{align*}
\bbE[\cL_{VL}(\bs)]\ge \frac{1}{360}BL\parenv{1  +o(1)}, \quad \text{as } B\rightarrow\infty,
\end{align*}
where the $o(1)$ term is independent of $\delta$.
\end{restatable}

\begin{IEEEproof}
Let $q=\min\parenv{2^{M-5},L/2}$. We find a set of distinct $M$-RLL $q$-substrings of $\bs$ that are encoded in full. In other words, any two such $q$-substrings are contained in two distinct chunks, or in two chunks that are duplicates, or in a single chunk without overlapping with each other. The total length of these $q$-substrings thus provides a lower bound on $\cL_{VL}(\bs)$.

Let $L_1,\ldots, L_A$ be given and assume $\cE_l$ holds. We consider the first $\ceil{B/(4A)}$ descendants of each source symbol. Let $G_a$ denote the set of the first $\ceil{B/(4A)}$ descendants of $\gx_a$. Let $Q_a$ be the set containing all non-overlapping $q$-substrings of $G_a$, i.e., $Q_a=\{\bx_{1+(c-1)q,q}: \bx\in G_a, 1\le c\le c_a\}$, where $c_a=\floor{L_a/q}$ and let $Q=\cup_{a=1}^A Q_a$. For  $\bw\in \Sigma^q$, let $\bw\in Q$ denote the event that one of the substrings in $Q$ equals $\bw$. 
Applying Lemma~\ref{lem:wappearingrid} on $Q$ (with substring length equal to descendant length) yields 
\begin{align*}
    \Pr\parenv{\bw\in Q} & \ge \frac{1}{2}\min\parenv{1,\frac{1}{2}\parenv{\sum_{a=1}^A c_a}\frac{\cS_{\delta}\parenv{q,\ceil{\frac{B}{4A}}}}{2^q}} \\
    & =  \frac{1}{4}\ceil{\frac{B}{4A}}\frac{\sum_{a=1}^Ac_a}{2^q},
\end{align*}
where the equality follows from $q=\Omega(L)$ and the property that $\cS_{\delta}(\ell,m) = m$ if $m(1-\delta)^{\ell}\le 1$.
So the expected number of distinct $M$-RLL strings in $Q$ is at least
\begin{align*} 
\sum_{\bw\in R_M^q}\frac{1}{4}\ceil{\frac{B}{4A}}\frac{\sum_{a=1}^Ac_a}{2^q} &\ge \frac{1}{4}\parenv{2-\frac{1}{2^{M-2}}}^q\ceil{\frac{B}{4A}}\frac{\sum_{a=1}^Ac_a}{2^q}
    \\
    & \ge  \frac{1}{5}\cdot\ceil{\frac{B}{4A}}\sum_{a=1}^A c_a,
\end{align*}
for all $M>5$. Since the size of $Q$ is $\ceil{\frac{B}{4A}}\sum_{a=1}^A c_a$, by the Markov bound, with probability at least $\frac{1}{9}$, the number of distinct $M$-RLL $q$-strings in $Q$ is at least $\frac{1}{10} \ceil{\frac{B}{4A}}\sum_{a=1}^A c_a$.

Let $q'=\ceil{q/2}$. Consider the $q'$-substrings of source blocks $Y_1,\ldots,Y_B$, i.e., $(Y_b)_{i,q'}$ for all $b\in[B], i\in[\abs{Y_b}]$. Define $\cE_d$ to be the following event: for every two source blocks $Y_{b_1}$ and $Y_{b_2}$, the substring of $Y_{b_1}$ starting at position $i_1$ is different from the substring of $Y_{b_2}$ starting at position $i_2$, i.e.,  $(Y_{b_1})_{i_1,q'} \ne (Y_{b_2})_{i_2,q'}$, as long as $J_{b_1}\ne J_{b_2}$ or $i_1\ne i_2$. Since there are at most $(2BL)^2$ pairs of such substrings, by the union bound and Lemma~\ref{lem:idsstrins}, $\cE_d$ holds with probability at least
\begin{align*}
    1-(2BL)^2/{2^{q'}}.
\end{align*}

When $\cE_d$ holds, the distinct $M$-RLL $q$-substrings in $Q$ are then non-overlapping substrings of the dictionary and it takes $q$-bits to encode each of them. To see this, we consider the first time such $q$-strings appear in the source string. Let $(Y_b)_{j,q}$ be one of the $M$-RLL strings in $Q$. Given $\cE_d$, the only possible substrings of $\bs$ that equal $(Y_{b})_{k,q}$ are $(Y_1)_{k,q},\ldots,(Y_B)_{k,q}$. Let $b'$ be the smallest integer such that $(Y_{b'})_{j,q} = (Y_b)_{j,q}$. By the $M$-RLL property, $(Y_{b'})_{j,q}$ must be fully contained in a chunk. Moreover, this chunk must be a new chunk by the minimality of $b'$ and is entered into the dictionary. Similarly, every distinct $M$-RLL $q$-substring corresponds to a $q$-substring in the dictionary. Since strings in $Q$ do not overlap with each other, the corresponding $q$-substrings in the dictionary also do not overlap, and each takes $q$ bits to store.

Combining the two arguments, with probability at least $\frac{1}{9} - \frac{(2BL)^2}{2^{q'}}$, there are $\frac{1}{10}\ceil{\frac{B}{4A}}\sum_{a=1}^Ac_a$ distinct non-overlapping RLL substrings of length $q$, and each needs $q$ bits to be encoded. It sums up to 
\begin{align*}
    q\cdot\frac{1}{10}\ceil{\frac{B}{4A}}\sum_{a=1}^Ac_a \ge \frac{B}{40A}\sum_{a=1}^A\parenv{L_a-q}
\end{align*}
bits. Therefore,
\begin{align*}
    \bbE\brckt{\cL_{VL}(\bs)|\cE_l} &\ge \parenv{\frac{1}{9} - \frac{(2BL)^2}{2^{q'}}} \frac{B}{40A}\sum_{a=1}^A\parenv{L-q} \\
    &\ge \frac{BL}{360} \parenv{1+o(1)}.
\end{align*}
The desired result thus follows from \eqref{eq:probeuel}.
\end{IEEEproof}


The above three theorems are summarized in Corollary~\ref{cor:vld} in the Results section to show that poorly choosing $M$  prevents efficient compression by VLD.



In the next theorem, we give our upper bound on $\bbE\brckt{\cL_{VL}(\bs)}$. We consider the case when $2^M$ is of order $\Theta\parenv{\log B}$ and show that variable-length deduplication achieves high compression ratios.
\vlubd

\begin{IEEEproof}
First, encoding the length $\abs{\bs}$ takes $2\log\abs{\bs} + 3\le 2\log (BL) + 5$ bits. We study next the encoding of chunks. We adopt the same strategy as \cite{niesen2019information}: dividing chunks into two categories, interior chunks and boundary chunks. Consider all chunks whose first symbols are in $Y_b$ (see Figure~\ref{fig:bdinck}). Some chunks depend on the values of the
neighboring source blocks $Y_{b-1}$ and $Y_{b+1}$, i.e., it is possible to alter the chunk by replacing $Y_{b-1}$ or $Y_{b+1}$ with other strings. We call these the `boundary' chunks of $Y_b$. Other chunks are independent of the values of the neighboring source blocks. We call these the `interior' chunks of $Y_b$. Denote the set of interior chunks in $\bs$ by $\cC^{\circ}(\bs)$. Note that we consider the first chunk and the last chunk of the whole data stream as boundary chunks. It is pointed out in~\cite{niesen2019information} that the number of boundary chunks is upper bounded by $3(B+1)$ and the expected total length of boundary chunks is upper bounded by $B2^{M+2}$.\footnote{Although in~\cite{niesen2019information}, the upper bounds are derived for source strings produced by an edit-free source, the same upper bounds hold when edits exist since every source block is still a Bernoulli(1/2) process by itself.} Therefore, encoding unique boundary chunks takes at most $3(B+1) + B2^{M+2}$ bits.

\begin{figure}[h]
    \centering
    \scalebox{0.6}{\begin{tikzpicture}[
		node distance = 0cm,
		seq/.style={minimum width=.3cm, minimum height=0.5cm, shape=rectangle,draw,font=\small,inner sep=0mm},
		nonoverlap/.style={pattern=vertical lines, pattern color=black},
		marker/.style={pattern = north east lines, pattern color = green},
		overlap/.style={pattern=north west lines, pattern color=red},
		mu/.style={},
		]
    \node (ybname)[draw=none] {};
    \node (yb)[seq, minimum width = 5cm, below = of ybname]{};
    \node (lmk)[seq, minimum width = 0.5cm, xshift = -1.3cm, marker, below = of ybname]{$0^M$};
    \node (rmk)[seq, minimum width = 0.5cm, xshift = 1.5cm, marker, below = of ybname]{$0^M$};
    \node (yb+1)[seq, minimum width = 5cm, right = of yb]{};
    \node (yb+1name)[draw = none, above = of yb+1]{};
    \node (mk)[seq, minimum width = 0.5cm, below = of yb+1name, marker]{$0^M$}; 
    \node (yb-1)[seq,minimum width = 3cm, left = of yb]{};
    \node (yb-1name)[draw=none, above = of yb-1]{};
    \node (llmk)[seq,minimum width = 0.5cm, below = of yb-1name,marker, xshift = 1.25cm]{$0^M$};
    
    \draw[decorate,decoration={brace,amplitude=5pt,raise=1pt,mirror},yshift=-1pt]   ([xshift = 0.25cm]llmk.south)--([xshift = 0.25cm]lmk.south) node [black,midway, yshift = -0.5cm] 
    {\footnotesize boundary};
    
    \draw[decorate,decoration={brace,amplitude=5pt,raise=1pt,mirror},yshift=-1pt]   ([xshift = 0.25cm]lmk.south)--([xshift = 0.25cm]rmk.south) node [black,midway, yshift = -0.5cm] 
    {\footnotesize interior};
    
    \draw[decorate,decoration={brace,amplitude=5pt,raise=1pt,mirror},yshift=-1pt] ([xshift = 0.25cm]rmk.south) -- ([xshift = 0.25cm]mk.south) node [black,midway, yshift = -0.5cm] 
    {\footnotesize boundary};
    
    \draw[decorate,decoration={brace,amplitude=5pt,raise=1pt},yshift=-1pt]   ([xshift = -1.5cm]yb-1.north)--([xshift = 1.5cm]yb-1.north) node [black,midway, yshift = 0.5cm] 
    {$Y_{b-1}$};
    
    \draw[decorate,decoration={brace,amplitude=5pt,raise=1pt},yshift=-1pt]   ([xshift = -2.5cm]yb.north)--([xshift = 2.5cm]yb.north) node [black,midway, yshift = 0.5cm] 
    {$Y_{b}$};
    \draw[decorate,decoration={brace,amplitude=5pt,raise=1pt},yshift=-1pt]   ([xshift = -2.5cm]yb+1.north)--([xshift = 2.5cm]yb+1.north) node [black,midway, yshift = 0.5cm] 
    {$Y_{b+1}$};
    
    \end{tikzpicture}}
    \caption{Occurrences of boundary chunks and interior chunks of $Y_b$ in variable-length chunking.}
    \label{fig:bdinck}
\end{figure}
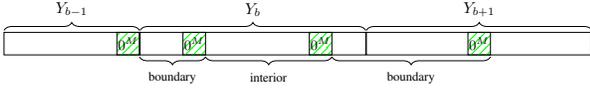

We consider next encoding unique interior chunks. Clearly, every interior chunk follows a $0^M$, i.e., the ending marker of the previous chunk. Moreover, this $0^M$ must also fully lie in the same source block as the chunk since otherwise this chunk is not an interior chunk. Therefore, the probability of occurrence of an interior chunk $\bu10^M$ is at most the probability of the occurrence of $0^M\bu10^M$ as a source block substring. It follows that
\begin{align}
    \bbE&\brckt{\sum_{\bw\in\cC^{\circ}(\bs)}\parenv{\abs{\bw}+1}|\cE_u} \nonumber\\
    \le & (M+1) + \sum_{\bu\in R_M} \Pr\parenv{\bu 10^M\in \cC^{\circ}(\bs)|\cE_u} \parenv{\abs{\bu}+M+2}\nonumber \\
    \le & (M+1) + \sum_{\bu\in R_M}\Pr\parenv{0^M\bu 10^M\in Y_1^B|\cE_u}\parenv{\abs{\bu}+M+2}, \label{eq:vlubddicbits}
\end{align}
where the term $M+1$ accounts for the chunk $0^M$. We compute the summation in \eqref{eq:vlubddicbits}. Fix $\gamma\in(\delta,1/2)$ and let $\ell_\gamma = \frac{\log (B/A)}{H(\gamma,\delta)}$.
\begin{itemize}
\item For all $\bu$ such that $\abs{0^M\bu10^M}\le \log B$, we trivially bound $\Pr\parenv{0^M\bu10^M\in Y_1^B|\cE_u}$ from above by 1. It follows that
\begin{multline}
    \sum_{\ell=0}^{\floor{\log B}-2M-1}\!\!\!\!\!\!\!\!\!\sum_{\bu\in R_M^\ell}\Pr\parenv{0^M\bu10^M\in Y_1^B|\cE_u}\parenv{\ell+M+2} \\
    \le \sum_{\ell=0}^{\floor{\log B}-2M-1}\sum_{\bu\in R_M^\ell}(\ell+M+2) \\
    \le \sum_{\ell=0}^{\floor{\log B}-2M-1}2^\ell\parenv{\ell+M+2}  \\
    \le \parenv{\floor{\log B}-M+1}2^{\log B-2M}  \\
    \le \frac{B\log B}{2^{2M}}.\label{eq:vlubddicbitssmall}
\end{multline}
\item For $\bu$ such that $\abs{0^M\bu 10^M} \ge \ell_\gamma$, we apply Lemma~\ref{lem:wappearlbdubd} and find \begin{align*}
    \Pr\parenv{0^M\bu10^M\in Y_1^B|\cE_u}&\le 2AL\frac{\cS_{\delta}\parenv{\abs{0^M\bu 10^M},\frac{3B}{2A}}}{2^{\abs{0^M\bu10^M}} }\\
    &\le \frac{3BL}{ 2^{\abs{0^M\bu 10^M}}}.
\end{align*}
It follows that 
\begin{align}
    &\quad\sum_{\mathclap{\ell=\ceil{\ell_\gamma}-2M-1}}^{2L}\ \parenv{\ell+M+2}\sum_{\bu\in R_M^\ell}\Pr\parenv{0^M\bu10^M\in Y_1^B|\cE_u}\nonumber\\
    &\le \sum_{\ell=\ceil{\ell_\gamma}-2M-1}^{2L}\sum_{\bu\in R_M^\ell}\frac{3BL}{2^{\ell+2M+1}} (\ell+M+2)\nonumber \\
    &\le \sum_{\ell=\ceil{\ell_\gamma}-2M-1}^{2L}2\parenv{2-\frac{1}{2^M}}^\ell \frac{3BL}{ 2^{\ell+2M+1}}\parenv{\ell+M+2}\nonumber\\
     &= \frac{3BL}{ 2^{2M}}\left(\sum_{\ell=\ceil{\ell_\gamma}-2M-1}^{2L}\parenv{1-\frac{1}{2^{M+1}}}^{\ell}(M+2)\right. \nonumber\\
    & \quad + \left.\sum_{\ell=\ceil{\ell_\gamma}-2M-1}^{2L}\parenv{1-\frac{1}{2^{M+1}}}^{\ell}\ell\right)\nonumber\\
    & = \parenv{1+o(1)}\frac{3BL}{ 2^{2M}}\left(2^{M+1}\cdot e^{-\frac{\ceil{\ell_{\gamma}}-2M-1}{2^{M+1}}}\right. \nonumber\\
    & \quad + \left. 2^{2(M+1)}\cdot e^{-\frac{\ceil{\ell_{\gamma}}-2M-1}{2^{M+1}}} \parenv{  \frac{\ceil{\ell_{\gamma}}-2M-1}{2^{M+1}}+1} \right)\nonumber\\
    &=  12BL\cdot e^{-\frac{\ell_{\gamma}}{2^{M+1}}}\parenv{\frac{\ell_\gamma}{2^{M+1}} + 1}\parenv{1+o(1)}\label{eq:vlubddicbitsmid},
\end{align}
where the second equality follows by applying summations \eqref{eq:sumbl} and \eqref{eq:sumlbl} in Appendix~\ref{app:summation} with $a=\ceil{\ell_{\gamma}}-2M-1$, $b = 2L$, $\beta = 2^{M+1}$ and noting that $\frac{2L}{2^{M+1}}=\omega(1)$.

\item If $\log B\le \ell_\gamma$, then there are additional terms corresponding to string $\bu$ such that $\log B\le\abs{0^M\bu10^M}\le \ell_\gamma$. Again by Lemma~\ref{lem:wappearlbdubd}, 
\begin{align*}
    \Pr\parenv{0^M\bu10^M\in Y_1^B|\cE_u}&\le 2AL\frac{\cS_{\delta}\parenv{\abs{0^M\bu 10^M},\frac{3B}{2A}}}{2^{\abs{0^M \bu 10^M}}}\\
    &\le 5BL 2^{-\abs{0^M\bu 10^M}(1+D(\gamma||\delta))},
\end{align*}
where the second inequality follows from \eqref{eq:cSdelta'} and the fact that $2^{n H(\gamma)} \le \frac{B}{A}2^{-n D\parenv{\gamma||\delta}}$ if $n\le \frac{\log(B/A)}{H(\gamma,\delta)}$.

Thus,
\begin{align}
    &\sum_{\ell=\ceil{\log B}-2M-1}^{\floor{\ell_\gamma}-2M-1}\sum_{\bu\in R_M^\ell} \Pr\parenv{0^M\bu10^M\in Y_1^B|\cE_u}\nonumber\\
    &\quad\cdot\parenv{\ell+M+2} \nonumber\\
    \le& \sum_{\ell=\ceil{\log B}-2M-1}^{\floor{\ell_\gamma}-2M-1}\sum_{\bu\in R_M^\ell} 5BL2^{-\parenv{\ell+2M+1}\parenv{1+D\parenv{\gamma||\delta}}} \nonumber\\&\quad\cdot\parenv{\ell+M+2}\nonumber\\
    \le & \frac{5BL}{2^{2M}}\!\!\!\sum_{\ell=\ceil{\log B}-2M-1}^{\floor{\ell_\gamma}-2M-1}\!\!\parenv{1-\frac{1}{2^{M+1}}}^{\ell}2^{-\parenv{\ell+2M+1}D\parenv{\gamma||\delta}}\nonumber\\
    &\quad \cdot\parenv{\ell+M+2}\nonumber\\
    \le &  \frac{5BL\ell_\gamma^2}{2^{2M}}\parenv{1-\frac{1}{2^{M+1}}}^{\log B-2M-1}2^{- D\parenv{\gamma||\delta}\log B}\nonumber\\
    = & \Theta\parenv{B^{1-D\parenv{\gamma||\delta}}L}\nonumber\\
    =&o(BL),\label{eq:vlubddicbitslg}
\end{align}
where the first equality follows from the fact that $\frac{\ell_{\gamma}^2}{2^{2M}}$ and $\parenv{1-\frac{1}{2^{M+1}}}^{\log B -2M-1}$ are both $\Theta(1)$ since $2^M$ and $\ell_\gamma$ are $\Theta\parenv{\log\parenv{B/A}}$.
\end{itemize}
Plugging \eqref{eq:vlubddicbitssmall}, \eqref{eq:vlubddicbitsmid} and \eqref{eq:vlubddicbitslg} in \eqref{eq:vlubddicbits}, we find that as $B\rightarrow \infty$ (also $A,L\rightarrow\infty$),
\begin{equation*}
    \bbE\brckt{\sum_{\bw\in \cC^{\circ}(\bs)}\parenv{\abs{\bw}+1}|\cE_u}\le 12e^{-c_M}\parenv{c_M+1}BL + o(BL),
\end{equation*}
where $c_M=\frac{\ell_\gamma}{2^{M+1}}$. 

If the complement of $\cE_u$ holds, then the number of bits needed for encoding interior chunks at their first occurrences is at most $4BL$, since the total length of interior chunks is at most $2BL$ and the total number of chunks is at most $2BL$. By noting that $\Pr\parenv{\bar\cE_u} \le Ae^{-\frac{B}{10A}}$,  
\begin{multline}
     \bbE\brckt{\sum_{\bw\in \cC^{\circ}(\bs)}\parenv{\abs{\bw}+1}} \\
    \le  12e^{-c_M}\parenv{c_M+1}BL+ o(BL) + 4BLAe^{-\frac{B}{10A}}\\
    =  12e^{-c_M}\parenv{c_M+1}BL\parenv{1+o(1)}. \label{eq:vlubdunqck}
\end{multline}

The number of bits needed for encoding pointers of repeated chunks can be bounded from above in a trivial way. Note that there are at most $\frac{\abs{\bs}}{M}+1$ strings in the dictionary $T$. So a pointer takes at most $\log\parenv{\frac{\abs{\bs}}{M}+1}+1\le\log \abs{\bs}$ bits. Moreover, the total number of chunks is less than the number of occurrences of $0^M$ plus 1 since every chunk except possibly the last one ends with $0^M$. On average, the number of occurrences of $0^M$ in $Y_b$ is at most $\frac{\abs{Y_b}}{2^M}$. So given $\abs{\bs}$, the expected number of chunks in $\bs$ is at most $\frac{\abs{\bs}}{2^M}+B+1$. Therefore the expected number of bits used by pointers is at most 
\begin{align}
    \bbE&\left[(\log\abs{\bs}+1)\cdot \parenv{\frac{\abs{\bs}}{2^M} + B + 1}  \right] \nonumber\\
    &\le  \log (2BL+1) \parenv{\frac{2BL}{2^M}+B+1}\nonumber\\
    &\le  2BL\frac{\log(BL)}{2^M} \parenv{1+o(1)}\nonumber\\
    &\le  4 H\parenv{\gamma,\delta} \frac{ \parenv{1+k_1}}{k_2} c_M \cdot BL\parenv{1+o(1)},
    \label{eq:pointer}
\end{align}
where the last inequality follows from $\frac{\log(BL)}{\log(B/A)} \le \frac{1+k_1}{k_2}\parenv{1+o(1)}$.

The desired result follows from summing  \eqref{eq:vlubdunqck} and \eqref{eq:pointer} and noting that the number of bits used for encoding the length of $\bs$ and the unique boundary chunks are $o(BL)$.
\end{IEEEproof}
A detailed analysis in the Results section shows that as $\delta$ approaches $0$, by appropriately choosing $M$, the compression ratio $\frac{\bbE\brckt{\abs{\bs}}}{\bbE\brckt{\cL_{VL}(\bs)}}$ can get arbitrarily large.

\section{Conclusion}\label{sec:con}
In this paper, we studied the performance of deduplication algorithms on data streams with approximate repeats, a situation that is common in practice. For simplicity, we modeled the process producing approximate repeats as independent bit-wise Bernoulli substitutions. We showed, in particular, that correctly choosing the chunk lengths is critical to the success of deduplication. 
With optimally chosen chunk lengths, deduplication in the fixed-length scheme is shown to achieve performance within a constant factor of optimal for a specific family of source models and with the knowledge of source  parameters. Additionally, appropriately choosing the length of the marker leads to suitable chunk lengths for variable-length deduplication, resulting in arbitrarily large compression ratios as source entropy gets smaller.

While this work sheds light on certain important aspects of the problem, the information-theoretic analysis of data deduplication provides a wealth of open problems. For example, while VLD was shown to achieve high compression ratios,  it is not known whether it is order optimal. Moreover, the source model proposed in this paper only included independent substitution edits. However, in practice, insertions, deletions and substitutions of single symbols, as well as longer strings, occur frequently. The probabilistic description of the source models can also be further refined based on experiments. Therefore, to gain a fuller understanding, it is important to study deduplication algorithms under  more general source models and edit processes.

\appendices
\section{Proof of Lemma~\ref{lem:RLL}}\label{app:numRLL} 
\lemRLL*
\begin{IEEEproof}
Clearly, if $0\le n\le k-1$, then any string of length $n$ is a $k$-RLL string (we consider the empty string as the only string of length 0). Therefore, for all $0\le n\le k-1$,
\begin{equation*}
     \abs{R_k^n} = 2^n \ge 
     (2-\frac{1}{2^{k-2}})^n,
\end{equation*}
and 
\begin{align*}
    \abs{R_k^n} &= 2^n = 2^{n+1} 2^{-1} \le 2^{n+1}(1-\frac{1}{2^{k+1}})^{k-1}
    \\ &\le 2^{n+1}(1-\frac{1}{2^{k+1}})^n = 2(2-\frac{1}{2^{k}})^n.
\end{align*}

For $n\ge k$, we prove the lemma by induction on $n$. Suppose the desired results hold for all $n'<n$. It is shown in~\cite[Chapter~8]{sedgewick2013introduction} that $\abs{R_k^N}=\sum_{i=1}^{k}\abs{R_k^{N-i}}$ for all $N\ge k$. Therefore,
\begin{align*}
    \abs{R_k^n} & = \sum_{i=1}^{k} \abs{R_k^{n-i}} \ge \sum_{i=1}^k (2-\frac{1}{2^{k-2}})^{n-i} \\
    &=  \frac{(2-\frac{1}{2^{k-2}})^n - (2-\frac{1}{2^{k-2}})^{n-k}}{1-\frac{1}{2^{k-2}}}\\
    & =  \parenv{2-\frac{1}{2^{k-2}}}^n \\
    &\quad + \frac{(2-\frac{1}{2^{k-2}})^{n-k}2^k}{2^{k-2}-1} \parenv{(1-\frac{1}{2^{k-1}})^k-\frac{1}{4}}
    \\&\ge  (2-\frac{1}{2^{k-2}})^n,
\end{align*}
and 
\begin{align*}
    \abs{R_k^n} &=\sum_{i=1}^k \abs{R_{k}^{n-i}} \le \sum_{i=1}^k 2(2-\frac{1}{2^k})^n 
    \\
    & =  2 \frac{(2-\frac{1}{2^k})^n - (2-\frac{1}{2^k})^{n-k}}{1-\frac{1}{2^k}} \\
    & =  2(2-\frac{1}{2^k})^n + \frac{2(2-\frac{1}{2^k})^{n-k}}{1-\frac{1}{2^k}}\parenv{\parenv{\frac{2-\frac{1}{2^k}}{2}}^k -1 }
    \\
    &\le 2(2-\frac{1}{2^k})^n.
\end{align*}
\end{IEEEproof}
\section{Proof of Lemma~\ref{lem:wapp1col}}\label{app:wapp1col}
\wapponecol*
\begin{IEEEproof}
We first prove inequality \eqref{eq:Prwinrm}. Given $\br$, the probability of a $\delta$-edit descendant being equal to $\bw$ is $\delta^{d_{\bw,\br}}(1-\delta)^{\ell-d_{\bw,\br}}$, where $d_{\bw,\br}$ denotes the Hamming distance between $\bw$ and $\br$. Therefore,
\begin{align*}
    \Pr\parenv{\bw\in\br_{[m]}} &= 1-\Pr\parenv{\bw\notin\br_{[m]}}\\
    & = 1-\sum_{\br\in \Sigma^\ell}\Pr(\br) \Pr(\bw\neq\br_1|\br)^m\\
    & = 1-\sum_{\br\in\Sigma^\ell}\Pr\parenv{\br}\parenv{1-\delta^{d_{\bw,\br}}(1-\delta)^{\ell-d_{\bw,\br}}}^m\\
    & = 1-\sum_{t=0}^{\ell}\parenv{\frac{{\ell\choose t}}{2^\ell}\parenv{1-\delta^t(1-\delta)^{\ell-t}}^m},
\end{align*}
where the second equality follows from the fact that $\br_1,\br_2,\ldots,\br_m$ are iid given $\br$ and the last equality follows from the fact that there are ${{\ell\choose t}}$ strings of length $\ell$ that are at Hamming distance $t$ from $\bw$. The desired inequalities then follow directly from applying inequalities \eqref{eq:1-xton} on $1-\parenv{1-\delta^t\parenv{1-\delta}^{\ell-t}}^m$.

The expected number of unique strings in $\br_{[m]}$ equals 
\begin{equation*}
    \bbE\brckt{\sum_{\bw\in\Sigma^{\ell}}I_{\bw\in\br_{[m]}}} = \sum_{\bw\in\Sigma^{\ell}} \Pr\parenv{\bw\in\br_{[m]} }. 
\end{equation*}
So the upper bound $\cS_{\delta}\parenv{\ell,m}$ and the lower bound $\frac{1}{2}\cS_{\delta}\parenv{\ell,m}$ follow from replacing $\Pr\parenv{ \bw\in\br_{[m]}}$ with its upper and lower bounds, respectively.

We show that $\cS_{\delta}\parenv{\ell,m}$ takes the given values for different $m$ and $\ell$:
\begin{itemize}
    \item When $\ell\ge \frac{\log m}{H(\delta)}$, $m\delta^{\delta\ell}\parenv{1-\delta}^{\parenv{1-\delta}\ell} \le 1$. It follows that 
    \begin{align*}
        \cS_{\delta}\parenv{\ell,m} &\ge \sum_{t=\ceil{\delta\ell}}^{\ell}{\ell\choose t}\min\parenv{1,m\delta^t(1-\delta)^{\ell-t}} \\
        &= \sum_{t=\ceil{\delta\ell}}^{\ell}{\ell\choose t}m\delta^t(1-\delta)^{\ell-t}\ge\frac{1}{4}m, 
    \end{align*}
    where the equality follows from the fact that $m\delta^t(1-\delta)^{\ell-t}$ is decreasing in $t$ so $m\delta^t\parenv{1-\delta}^{\ell-t}\le 1$ for all $t\ge \delta\ell$ and the second inequality follows from the result shown in~\cite{greenberg2014tight} that for a binomial random variable $X$ with parameters $n$ and $p$,~\(\Pr\parenv{X\ge np}>\frac{1}{4}\) if \(p\ge 1/n\). 
    
    Moreover, when $\ell \ge \frac{\log m}{\log(\frac{1}{1-\delta})}$, $m\delta^t\parenv{1-\delta}^{\ell-t}\le 1$ for all $t$. Hence, 
\begin{equation*}
    \cS_{\delta}\parenv{\ell,m} = \sum_{t=0}^{\ell}{\ell\choose t}m\delta^t(1-\delta)^{\ell-t} = m.
\end{equation*}

\item When $\ell\le \frac{\log m}{H(\frac{1}{2},\delta)}$, $m\delta^{\frac{\ell}{2}}\parenv{1-\delta}^{\frac{\ell}{2}} \ge 1$. It follows that
\begin{align*}
    \cS_{\delta}\parenv{\ell,m} &\ge \sum_{t=0}^{\floor{\frac{\ell}{2}}}{\ell\choose t}\min\parenv{1,m\delta^t(1-\delta)^{\ell-t}}= \sum_{t=0}^{\floor{\frac{\ell}{2}}} {\ell\choose t} \\
    &\ge 2^{\ell-1},
\end{align*}
where the first inequality follows from the fact that $m\delta^{t}\parenv{1-\delta}^{\ell-t} \ge 1$ for all $t\le \frac{\ell}{2}$.

Moreover, when $\ell\le \frac{\log m}{ \log(\frac{1}{\delta})}$, $m\delta^t \ge 1$ for all $t$. Hence,
\begin{equation*}
    \cS_{\delta}\parenv{\ell,m} = \sum_{t=0}^{\ell}{\ell\choose t}\cdot 1 = 2^\ell. 
\end{equation*}

\item For any $\delta<\delta'<1/2$,
\begin{align*}
    \cS_{\delta}\parenv{\ell,m}&\le \sum_{t=0}^{\floor{\delta'\ell}} \binom{\ell}{t} + \sum_{t=\ceil{\delta'\ell}}^{\ell} \binom{\ell}{t}m\delta^t(1-\delta)^{\ell-t}\\
        &\le 2^{\ell H(\delta')}+ m2^{-\ell D(\delta'||\delta)},
\end{align*}
where the second inequality follows from applying the Chernoff bound on a binomial distribution with parameters $\ell$ and $\delta$.

When $\ell = \frac{\log m}{H(\delta',\delta)}$, $2^{\ell H(\delta')} = m2^{-\ell D\parenv{\delta'||\delta}}$. So \(2^{\ell H(\delta')}+ m2^{-\ell D(\delta'||\delta)} = 2^{\ell H(\delta')+1}\) and
\begin{equation*}
    \cS_\delta\parenv{\ell,m}\le 2^{\ell H\parenv{\delta'}+1}.
\end{equation*}

\item The upper bounds $2^\ell$ and $m$ follow from:
\begin{align*}
    &\cS_{\delta}\parenv{\ell,m} \le \sum_{t=0}^\ell \frac{{\ell\choose t}}{2^\ell} = 1, \\
    &\cS_{\delta}\parenv{\ell,m} \le \sum_{t=0}^\ell \frac{{\ell\choose t}}{2^\ell}m\delta^t\parenv{1-\delta}^{\ell-t} = \frac{m}{2^\ell}.
\end{align*}
\end{itemize}
\end{IEEEproof}
\section{Proofs of Lemma~\ref{lem:wappearingrid} and Lemma~\ref{lem:dicsizefx}}\label{app:wappearingrid}\label{app:dicsizefx}
\lemwappearingrid*
\begin{IEEEproof}
Let the $K$ strings be denoted $\by_1,\by_2,\ldots,\by_K$. Let $G_\bw(i)$ denote the event that $\bw$ equals one of the descendants of $\by_i$. Clearly, $G_{\bw}(1),G_{\bw}(2),\ldots,G_\bw\parenv{K}$ are independent and
\begin{align}
    G_\bw = \cup_{i=1}^K G_\bw(i).\label{eq:unioniGbw}
\end{align}
Note that by Lemma~\ref{lem:wapp1col} and the fact that $\cS_\delta\parenv{n,m}$ is non-decreasing in $m$,
\begin{align*}
    \frac{1}{2}\frac{\cS_\delta\parenv{n,m_1}}{2^{n}}\le \Pr\parenv{G_\bw(i)} \le  \frac{\cS_\delta\parenv{n,m_2}}{2^{n}}.
\end{align*}

Applying the union bound on \eqref{eq:unioniGbw} gives
\begin{align*}
    \Pr\parenv{G_\bw} \le \sum_{i=1}^K\Pr\parenv{G_\bw(i)} \le K\frac{\cS_\delta(n,m_2)}{2^{n}}.
\end{align*}
The desired upper bound follows by noting that 1 is a trivial upper bound.

We then prove the lower bound. By independence, 
\begin{align*}
   \Pr\parenv{G_\bw}&=\Pr\parenv{\cup_{i=1}^K G_\bw(i)} \\
   &= 1-\prod_{i=1}^K  \parenv{1-\Pr\parenv{G_\bw\parenv{i}}}\\
   & \ge 1-\parenv{1-\frac{1}{2}\frac{\cS_\delta\parenv{n,m_1}}{2^{n}}}^{K}\\
   &\ge \frac{1}{2}\min\parenv{1,\frac{1}{2}K\frac{\cS_\delta\parenv{n,m_1}}{2^{n}}},
\end{align*}
where the last inequality follows from inequality \eqref{eq:1-xton} that $1-(1-x)^n \ge \frac{1}{2}\min\parenv{1,nx}$ for $x\in\parenv{0,1}$ and integer $n$.
\end{IEEEproof}
\corwappearingrid*
\begin{IEEEproof}
The size of $T_F^1(\bs)$ equals the number of distinct strings among chunks $Z_c^b,1\le c \le C+1,1\le b\le B$. Clearly, chunks of length $\ell$ are $\delta$-edit descendants of the $AC$ source symbol substrings $U_c^a, 1\le c\le C,1\le a\le A$, which are independent and uniformly distributed in $\Sigma^\ell$. Given $\cE_u$, each $U_c^a$ has at most $\frac{3B}{2A}$ descendants. Moreover, since we assume that the source symbols $\gx_1,\ldots,\gx_A$ are chosen uniformly and independently, it follows directly from Lemma~\ref{lem:wappearingrid} that for any $\ell$-string $\bw$,
\begin{equation*}
    \Pr\parenv{\bw\in T_F^1(\bs)|\cE_u} \le \min\parenv{1,AC\frac{\cS_\delta\parenv{\ell,\frac{3B}{2A}}}{2^\ell}}.
\end{equation*}
Hence
\begin{align*}
    \bbE\brckt{\abs{T_F^1(\bs)}|\cE_u} &\le  \sum_{\bw\in\Sigma^\ell} \Pr\parenv{\bw\in T_F^1(\bs)|\cE_u} + B \\
    &\le \min\parenv{2^\ell,AC\cS_\delta\parenv{\ell,\frac{3B}{2A}}} +B,
\end{align*}
where the addend $B$ accounts for the chunks of lengths less than $\ell$ at the end of each source block, if any.

The lower bound on $\abs{T_F^{\nicefrac{1}{2}}(\bs)}$ given $\cE_l$ follows similarly from Lemma~\ref{lem:wappearingrid}.
\end{IEEEproof}

\section{Proof of Lemma~\ref{lem:wappearingridsub} and Lemma~\ref{lem:wappearlbdubd}}\label{app:wappearingridsub}\label{app:wappearlbdubd}
\lemwappearingridsub*
\begin{IEEEproof}
Let the $K$ strings be denoted $\by_1,\by_2,\ldots,\by_K$. We use $\cD_i$ to denote the set of $\delta$-edit descendants of $\by_i$. Let $H_\bw(i,j)$ denote the event that $\bw=\bx_{j,\abs{\bw}}$ for some $\bx\in\cD_i$. Clearly, 
\begin{align}
    H_\bw = \cup_{i=1}^K\cup_{j=1}^{n-\abs{\bw}+1}H_\bw(i,j).\label{eq:unionHbw}
\end{align}
Note that the strings $\{\bx_{j,\abs{\bw}}\}_{\bx\in\cD_i}$ are iid $\delta$-edit descendants of $(\by_i)_{j,\abs{\bw}}$. Hence by Lemma~\ref{lem:wapp1col}
\begin{multline*}
    \frac{1}{2}\frac{\cS_\delta\parenv{\abs{\bw},m_1}}{2^{\abs{\bw}}}\le\frac{1}{2}\frac{\cS_\delta\parenv{\abs{\bw},\abs{\cD_i}}}{2^{\abs{\bw}}}\\
    \le \Pr\parenv{H_\bw(i,j)} \le \\\frac{\cS_\delta\parenv{\abs{\bw},\abs{\cD_i}}}{2^{\abs{\bw}}}\le \frac{\cS_\delta\parenv{\abs{\bw},m_2}}{2^{\abs{\bw}}}.
\end{multline*}
where the first and the last inequalities follow from $m_1\le \abs{\cD_i}\le m_2$.

Applying the union bound on \eqref{eq:unionHbw} gives
\begin{align*}
    \Pr\parenv{H_\bw} & \le \cup_{i=1}^K\cup_{j=1}^{n-\abs{\bw}+1}\Pr\parenv{H_\bw(i,j)} \\
    &\le \parenv{n-\abs{\bw}+1}K\frac{\cS_\delta(\abs{\bw},m_2)}{2^{\abs{\bw}}}.
\end{align*}
The desired upper bound follows by noting that 1 is a trivial upper bound.


We next prove the lower bound. For each $i$, non-overlapping substrings of $\br_i$ are independent and so are their descendants. Hence, events $H_\bw(i,j)$, $j=1,1+\abs{\bw},\ldots,1+(p-1)\abs{\bw}$, where $p=\floor{\frac{n}{\abs{\bw}}}$, are mutually independent. It follows that
\begin{align*}
   &\quad \Pr\parenv{\cup_{i=1}^K\cup_{a=1}^p H_\bw(i,1+(a-1)\abs{\bw})} \\
   &= 1-\prod_{i=1}^K \prod_{a=1}^p \parenv{1-\Pr\parenv{H_\bw\parenv{i,1+(a-1)\abs{\bw}}}}\\
   & \ge 1-\parenv{1-\frac{1}{2}\frac{\cS_\delta\parenv{\abs{\bw},m_1}}{2^{\abs{\bw}}}}^{Kp}\\
   &\ge \frac{1}{2}\min\parenv{1,\frac{1}{2}Kp\frac{\cS_\delta\parenv{\abs{\bw},m_1}}{2^{\abs{\bw}}}},
\end{align*}
where the last inequality follows from inequality \eqref{eq:1-xton} that $1-(1-x)^n \ge \frac{1}{2}\min\parenv{1,nx}$ for $x\in\parenv{0,1}$ and integer $n$. The desired lower bound thus follows by noting that 
\begin{align*}
    \cup_{i=1}^K\cup_{a=1}^p H_\bw(i,1+(a-1)\abs{\bw}) \subseteq H_\bw.
\end{align*}
\end{IEEEproof}

\wappearlbdubd*
\begin{IEEEproof}
Recall that we assume every source symbol (and thus every source block) is of length at least $\frac{1}{2}L $ and at most $ 2L$. So we can get a lower bound on $\Pr\parenv{\bw\in Y_1^{B/2}|\cE_l}$ by assuming every source block is of length $\frac{L}{2}$. Similarly, we get an upper bound on $\Pr\parenv{\bw\in Y_1^B|\cE_u}$ by assuming every source block is of length $2L$. 

Now that the $B$ source blocks are independent and each is a $\delta$-edit descendant of one of the $A$ source symbols. Moreover, each random string (source symbol) has at most $\frac{3B}{2A}$ descendants given $\cE_u$. Therefore, by directly applying Lemma~\ref{lem:wappearingridsub},
\begin{align*}
     \Pr\parenv{\bw\in Y_1^B|\cE_u} &\le \min\parenv{1,\parenv{2L-\abs{\bw}+1}A\frac{\cS_\delta\parenv{\abs{\bw},\frac{3B}{2A}}}{2^{\abs{\bw}}}} \\
     &\le \min\parenv{1,2LA\frac{\cS_\delta\parenv{\abs{\bw},\frac{3B}{2A}}}{2^{\abs{\bw}}}}.
\end{align*}
The lower bound can be obtained similarly:
\begin{align*}
    \Pr\parenv{\bw\in Y_1^{B/2}|\cE_l} &\ge \frac{1}{2}\min\parenv{1,\frac{1}{2}\floor{\frac{L/2}{\abs{\bw}}}A\frac{\cS_\delta\parenv{\abs{\bw},\frac{B}{4A}}}{2^{\abs{\bw}}}} \\
    &\ge \frac{1}{2}\min\parenv{1,\frac{1}{8}\frac{L}{\abs{\bw}}A\frac{\cS_\delta\parenv{\abs{\bw},\frac{B}{4A}}}{2^{\abs{\bw}}}}.
\end{align*}
\end{IEEEproof}

\section{Proof of Lemma~\ref{lem:wappearlbdbigellvl}}\label{app:wappearlbdbigellvl}
\wappearlbdbigellvl*
\begin{IEEEproof}
Let $\bw=10^M\bu 10^M$. By assumption, $\abs{\bw} = \abs{\bu}+2M+2 \ge \frac{\log(B/A)-2}{H(\delta)}$. 

For definiteness, we assume $\abs{Y_{\nicefrac{1}{2}}(a)}=\frac{B}{4A}$ for all $a$ and all source symbols are of length $\frac{L}{2}$. With these assumptions, we have a similar setting to that in Lemma~\ref{lem:wappearingridsub}. So we adopt the same notation. Let $H_\bw$ denote $\bw\in Y_1^{B/2}$ and $H_\bw(a,j)$ denote the event that $\bw=\bx_{j,\abs{\bw}}$ for some $\bx\in Y_{\nicefrac{1}{2}}(a)$. Similar to \eqref{eq:unionHbw}:
\begin{align}
    H_\bw = \cup_{a=1}^A\cup_{j=1}^{\ceil{L/2}-\abs{\bw}+1}H_\bw\parenv{a,j}.\label{eq:unionAj}
\end{align}
Moreover, 
\begin{align*}
    \frac{1}{2}\frac{\cS_\delta\parenv{\abs{\bw},\frac{B}{4A}}}{2^{\abs{\bw}}} \le \Pr\parenv{H_\bw\parenv{a,j}} \le \frac{\cS_\delta\parenv{\abs{\bw},\frac{B}{4A}}}{2^{\abs{\bw}}}.
\end{align*}

In Lemma~\ref{lem:wappearlbdubd}, an upper bound on $\Pr\parenv{\bw\in Y_1^{B/2}|\cE_l}$ is obtained by applying the union bound on \eqref{eq:unionAj}. Here, we get a lower bound by the inclusion-exclusion principle:
\begin{align}
    &\Pr\parenv{H_\bw} \ge \sum_{a=1}^A\sum_{i=1}^{\ceil{L/2}-\abs{\bw}}\Pr\parenv{H_\bw\parenv{a,i}} \label{eq:wapplbdunbd1}\\
    & - \!\! \sum_{1\le a_1\ne a_2\le A}\!\!\sum_{j=1}^{\ceil{L/2}-\abs{\bw}}\sum_{k=1}^{\ceil{L/2}-\abs{\bw}}\!\!\Pr\parenv{H_\bw\parenv{a_1,j}\cap H_\bw\parenv{a_2,k}} \label{eq:wapplbdunbd2}\\
    & -\sum_{a=1}^A \;\;\sum_{\substack{1\le j,k \le \ceil{L/2}-\abs{\bw}\\ j\ne k}} \Pr\parenv{H_\bw\parenv{a,j}\cap H_\bw\parenv{a,k}}.\label{eq:wapplbdunbd3}
\end{align}
We compute the three terms on the right-hand side of the inequality above as follows.

For the term in \eqref{eq:wapplbdunbd1}, since $\abs{\bw}\ge \frac{\log(B/A)-2}{H(\delta)}$, 
\begin{align*}
    \Pr\parenv{H_\bw\parenv{a,i}}&\ge \frac{1}{2}\frac{\cS_{\delta}\parenv{\abs{\bw},\ceil{\frac{B}{4A}}}}{2^{\abs{\bw}}} \ge \frac{1}{2} \frac{\cS_{\delta}\parenv{\abs{\bw},\frac{B}{4A}}}{2^{\abs{\bw}}}\\
    &\ge  \frac{B}{32A\cdot 2^{\abs{\bw}}},
\end{align*}
where the last inequality follows from \eqref{eq:sellmlargeell}. It follows that
\begin{align}
    \sum_{a=1}^A\sum_{i=1}^{\ceil{L/2}-\abs{\bw}}\Pr\parenv{H_\bw\parenv{a,i}} &\ge A\parenv{\ceil{L/2}-\abs{\bw}}\frac{B}{32A\cdot 2^{\abs{\bw}}} \nonumber\\
    &\ge\frac{BL}{2^7\cdot 2^{\abs{\bw}}}.\label{eq:wapplbdin1}
\end{align}

For the term in \eqref{eq:wapplbdunbd2}, since for all $a_1\ne a_2$, $\by_{a_1}$ and $\by_{a_2}$ are independent and so are their descendants, we get
\begin{align}
    &\sum_{1\le a_1\ne a_2\le A}\sum_{j=1}^{\ceil{L/2}-\abs{\bw}}\sum_{k=1}^{\ceil{L/2}-\abs{\bw}}\Pr\parenv{H_\bw\parenv{a_1,j}\cap H_\bw\parenv{a_2,k}}\nonumber\\
    &=\!\! \sum_{1\le a_1\ne a_2\le A}\!\!\sum_{j=1}^{\ceil{L/2}-\abs{\bw}}\!\!\sum_{k=1}^{\ceil{L/2}-\abs{\bw}}\!\!\!\Pr\parenv{H_\bw\parenv{a_1,j}}\Pr\parenv{H_\bw\parenv{a_2,k}}\nonumber\\
    &\le \sum_{1\le a_1\ne a_2\le A}\sum_{j=1}^{\ceil{L/2}-\abs{\bw}}\sum_{k=1}^{\ceil{L/2}-\abs{\bw}}\parenv{\frac{\cS_{\delta}\parenv{\abs{\bw},\ceil{\frac{B}{4A}}}}{2^{\abs{\bw}}}}^2\nonumber\\
    & \le \sum_{1\le a_1\ne a_1\le A}\frac{B^2L^2}{ A^2 2^{2\abs{\bw}}}\nonumber\\
    & \le \frac{B^2L^2}{ 2^{2\abs{\bw}}},  \label{eq:wapplbdin2}
\end{align}
where the second inequality follows from \eqref{eq:sellmubd} that $\cS_\delta\parenv{\ell,m}\le m$ and the inequalities $\ceil{L/2}-\abs{\bw}\le L$, $\ceil{\frac{B}{4A}}\le \frac{B}{A}$.

We then consider the term in \eqref{eq:wapplbdunbd3}, where the two occurrences of $\bw$ are among the descendants of a single source symbol, and thus might not be independent. Unlike the previous two terms, we consider lower bounding the sum of probabilities $\Pr\parenv{H_\bw\parenv{a,j}\cap H_\bw\parenv{a,k}}$ over all $\bw$ of the form $10^M\bu 10^M,\bu\in R_M^n$. For clarity of presentation, we first claim (to be proved later) that for any $a$,
\begin{align}
    &\sum_{\substack{\bw:\bw=10^M\bu10^M\\ \bu\in R_M^n}}\;\sum_{\substack{1\le j, k \le \ceil{L/2}-\abs{\bw}\\j\ne k}} \Pr\parenv{H_\bw\parenv{a,j}\cap H_\bw\parenv{a,k}}\nonumber
    \\
    &\le\frac{B^2L^2}{A^22^{\abs{\bw}}}\parenv{1 + \frac{n+M+1}{L} }.  \label{eq:appenB}
\end{align}
It follows that 
\begin{align}
    &\sum_{\substack{\bw:\bw=10^M\bu10^M\\ \bu\in R_M^n}}\;\sum_{a=1}^A\;\sum_{\substack{1\le j, k \le \ceil{L/2}-\abs{\bw}\\j\ne k}} \!\!\!\Pr\parenv{H_\bw\parenv{a,j}\cap H_\bw\parenv{a,k}} \nonumber
    \\
    &\le \frac{B^2L^2}{A 2^{\abs{\bw}}}\parenv{1+\frac{n+M+1}{L}}.\label{eq:wapplbdin3w}
\end{align}

Thus, combining \eqref{eq:wapplbdunbd3}, \eqref{eq:wapplbdin1}, \eqref{eq:wapplbdin2} and \eqref{eq:wapplbdin3w} gives 
\begin{align*}
    &\sum_{\substack{\bw:\bw=10^M\bu10^M\\ \bu\in R_M^n}}\Pr\parenv{H_\bw} \\
    &\ge \sum_{\substack{\bw:\bw=10^M\bu10^M\\ \bu\in R_M^n}}
    \parenv{\frac{BL}{2^7\cdot 2^{\abs{\bw}}}  - \frac{B^2 L^2}{2^{2\abs{\bw}}}} \\&\quad -\frac{B^2L^2}{A 2^{\abs{\bw}}}\parenv{1+\frac{n+M+1}{L}}\\
    &\ge\frac{BL}{2^7\cdot 2^{\abs{\bw}}}\cdot\abs{R_M^{n}} \\
    &\quad - \frac{B^2 L^2}{2^{2\abs{\bw}}}\cdot \abs{R_M^{n}} -\frac{B^2L^2}{A 2^{\abs{\bw}}}\parenv{1+\frac{n+M+1}{L}}\\
    &\ge \frac{BL}{2^7\cdot 2^{\abs{\bw}}}\cdot\abs{R_M^{n}} - \frac{3B^2L^2}{2^{\abs{\bw}}},
\end{align*}
where the last inequality follows from $\abs{R_M^n} \le 2^{\abs{\bw}}$, $\frac{n+M+1}{L} \le 1$ and $A\ge 1$. The desired lower bound thus follows from bounding $\abs{R_M^n}$ by Lemma~\ref{lem:RLL}.

Finally, we prove inequality \eqref{eq:appenB}. Fix $a$. That $H_\bw\parenv{a,j}$ and $H_\bw\parenv{a,k}$ both hold means there exist descendants $\bx_1,\bx_2$ (possibly the same one) of $\by_a$ such that $(\bx_1)_{j,\abs{\bw}}=(\bx_2)_{k,\abs{\bw}}=\bw$. Assume $j< k$ without loss of generality. We compute $\Pr\parenv{H_\bw\parenv{a,j}\cap H_\bw\parenv{a,k}}$ for different values of $(j,k)$: 
\begin{itemize}
    \item $\abs{j-k}\ge\abs{\bw}$. The two occurrences of $\bw$ in $\bx_1$ and $\bx_2$ are plotted in Figure~\ref{fig:twooccurwcase1}. In this case, they are produced by two non-overlapping substrings of $\by_a$ and thus are independent. It follows that 
\begin{align}
      &\sum_{\abs{j-k}\ge\abs{\bw}} \Pr\parenv{H_\bw\parenv{a,j}\cap H_\bw\parenv{a,k}} \nonumber\\
      &= \sum_{\abs{j-k}\ge\abs{\bw}} \Pr\parenv{H_\bw\parenv{a,j}}\Pr\parenv{H_\bw\parenv{a,k}}\nonumber\\
      &\le L^2\parenv{\frac{\cS_{\delta}\parenv{\abs{\bw},\ceil{\frac{B}{4A}}}}{2^{\abs{\bw}}}}^2\nonumber\\
      & \le \frac{L^2B^2}{A^2 2^{2\abs{\bw}}}. \label{eq:wapplbdin31}
\end{align}
\begin{figure}[h]
	\begin{center}
		\scalebox{0.65}{\begin{tikzpicture}[
		node distance = 0cm,
		seq/.style={minimum width=.3cm, minimum height=0.5cm, shape=rectangle,draw,font=\small,inner sep=0mm},
		marker/.style={pattern=dots, pattern color=black},
		nonoverlap/.style={pattern = north east lines, pattern color = green},
		overlap/.style={pattern=north west lines, pattern color=red},
		mu/.style={},
		]
		
		\node (root) [seq,minimum width=12cm] {};
		\node (rootname) [left = of root, xshift =-.3cm,seq, draw=none] {$\by_a$};

		\node (x1) [below = of root,yshift=-.5cm, seq,minimum width=12cm] {};
		\node (x1name) [left = of x1, xshift = -.3cm, seq, draw=none] {$\bx_1$};
		\node (w1lmk) [right = of x1name,xshift = 1cm,seq,minimum width = 0.8 cm]{$10^M$};
		\node (x1u) [right = of w1lmk, seq, overlap, minimum width = 3.4cm]{$\bu$};
		\node (w1rmk) [right = of x1u, seq, minimum width = 0.8cm]{$10^M$};

		\node (x2) [below = of x1,yshift=-.5cm, seq,minimum width=12cm] {};	
		\node (x2name) [left = of x2,xshift =-.3cm]{$\bx_2$};
        \node (w2lmk) [right = of x2name, xshift = 6.5cm, seq, minimum width =0.8cm]{$10^M$};
        \node (x2u) [right = of w2lmk, seq, overlap, minimum width = 3.4cm]{$\bu$};
        \node (w2rmk) [right = of x2u, seq, minimum width =0.8cm]{$10^M$};
        \draw [dashed] ([xshift = -0.4cm, yshift = 2cm]w1lmk.south) -- ([xshift = -0.4cm,yshift = 0cm]w1lmk.south);
        \node [above = of w1lmk,xshift = -0.4cm,yshift = 1.5cm ]{$j$};
        
        \draw [dashed] ([xshift = -0.4cm, yshift = 3cm]w2lmk.south) -- ([xshift = -0.4cm,yshift = 0cm]w2lmk.south);
        \node [above = of w2lmk,xshift = -0.4cm,yshift = 2.5cm ]{$k$};
        
        \draw[decorate,decoration={brace,amplitude=5pt,raise=1pt},yshift=-1pt]   ([xshift = -0.4cm]w1lmk.north)--([xshift = 0.4cm]w1rmk.north) node [black,midway, yshift = 0.35cm] 
    {$\bw$};
    
    \draw[decorate,decoration={brace,amplitude=5pt,raise=1pt},yshift=-1pt]   ([xshift = -0.4cm]w2lmk.north)--([xshift = 0.4cm]w2rmk.north) node [black,midway, yshift = 0.35cm] 
    {$\bw$};
		\end{tikzpicture}}
	\end{center}
	\caption{Relative position of the two occurrences of $\bw$ at position $j$ and $k$ when $\abs{j-k}\ge\abs{\bw}$.}
	\label{fig:twooccurwcase1}
\end{figure}

    \item $1\le\abs{j-k}<\abs{\bu}$. The two occurrences of $\bw$ in $\bx_1$ and $\bx_2$ are plotted in Figure~\ref{fig:twooccurwcase2}. In this case, the two occurrences of $\bw$ are descendants of two overlapping substrings of $\by_a$. Recall that $\bw=10^M\bu10^M$. We write the string $\bu$ in $\bx_1$ as $\bu_1\bu_2$, and write the string $\bu$ in $\bx_2$ as $\bu_2'\bu_3$, so that $\bu_2$ and $\bu_2'$ have the same ancestors, denoted $\br_2$. Denote the ancestor of $\bu_1$ and $\bu_3$ by $\br_1$ and $\br_3$, respectively. Denote the ancestor of $10^M$ at the beginning of $\bw$ in $\bx_1$ by $\br_0$, and the ancestor of $10^M$ at the end of $\bw$ in $\bx_2$ by $\br_4$. We have~$\abs{\br_1} = \abs{\bu_1} = \abs{\br_3}=\abs{\bu_3} = k-j$,~$\abs{\br_2} = \abs{\bu_2} = \abs{\bu_2'}=\abs{\bu}-(k-j)$. Write $\br=\br_0\br_1\br_2\br_3\br_4$.
    
    \begin{figure}[h]
	\centering
		\scalebox{0.65}{\begin{tikzpicture}[
		node distance = 0cm,
		seq/.style={minimum width=.3cm, minimum height=0.5cm, shape=rectangle,draw,font=\small,inner sep=0mm},
		marker/.style={pattern=dots, pattern color=black},
		nonoverlap/.style={pattern = north east lines, pattern color = green},
		overlap/.style={pattern=north west lines, pattern color=red},
		mu/.style={},
		]
		\node (rootname) [seq, draw=none] {$\by_a$};
		\node (root) [right = of rootname, xshift=.3cm, seq,minimum width=12cm] {};
	    \node (r0) [right = of rootname, xshift = 1cm,seq,minimum width = 0.8cm]{$\br_0$};
		\node (r1) [right = of rootname, xshift = 1.8cm,seq,nonoverlap,minimum width = 1.1cm]{$\br_1$};
		\node (r2) [right = of rootname, xshift = 2.9cm, seq,overlap, minimum width = 2.3cm]{$\br_2$};
        \node (r3) [right = of rootname, xshift = 5.2cm, seq, nonoverlap, minimum width = 1.1cm]{$\br_3$};
		\node (r4) [right = of rootname, xshift = 6.3cm, seq, minimum width = 0.8cm]{$\br_4$};

		\node (w1name) [below = of rootname,yshift=-.5cm, seq, draw=none] {$\bx_1$};
		\node (x1) [below = of root,yshift=-.5cm, seq,minimum width=12cm] {};
		
		\node (w1lmk) [below = of r0, yshift =-0.5cm,seq,minimum width = 0.8 cm]{$10^M$};
		\node (w1nonoverlap) [below = of r1, yshift = -0.5cm, seq, nonoverlap, minimum width = 1.1cm]{$\bu_1$};
		\node (w1overlap) [below = of r2,yshift=-0.5cm, seq, overlap, minimum width = 2.3cm]{$\bu_2$};
		\node (w1rmk) [right = of w1overlap, seq, minimum width = 0.8cm]{$10^M$};

		\node (w2name) [below =of w1name, yshift=-.5cm]{$\bx_2$};
		\node (x2) [below = of x1,yshift=-.5cm, seq,minimum width=12cm] {};
		\node (w2overlap) [below = of w1overlap, yshift = -.5cm, seq, overlap,minimum width = 2.3cm]{$\bu_2'$};
		\node (w2nonoverlap) [right = of w2overlap, seq, nonoverlap, minimum width = 1.1cm]{$\bu_3$};
        \node (w2rmk) [right = of w2nonoverlap, seq, minimum width =0.8cm]{$10^M$};
        \node (w2lmk) [left = of w2overlap, seq, minimum width = 0.8cm]{$10^M$};
        
        \draw [dashed] ([xshift = -0.4cm]w1lmk.south) -- ([xshift = -0.4cm,yshift = 2cm]w1lmk.south);
        \node [above = of r0,xshift = -0.4cm,yshift = .5cm ]{$j$};
        
        \draw [dashed] ([xshift = -0.4cm, yshift = 3cm] w2lmk.south) -- ([xshift = -0.4cm] w2lmk.south);
        \node [above = of w2lmk, yshift = 2.5cm, xshift = -0.4cm] {$k$};
        
        \draw[decorate,decoration={brace,amplitude=5pt,raise=1pt},yshift=-1pt]   ([xshift = -0.4cm]w1lmk.north)--([xshift = 0.4cm]w1rmk.north) node [black,midway, yshift = 0.35cm] 
    {$\bw$};
    \draw[decorate,decoration={brace,amplitude=5pt,raise=1pt},yshift=-1pt]   ([xshift = -0.4cm]w2lmk.north)--([xshift = 0.4cm]w2rmk.north) node [black,midway, yshift = 0.35cm] 
    {$\bw$};
		\end{tikzpicture}}
	\caption{Relative position of the two occurrences of $\bw$ at position $j$ and $k$ when $1\le\abs{j-k}<\abs{\bu}$.}
	\label{fig:twooccurwcase2}
	\end{figure}

For a single descendant $\bx$ of $\by_a$, $\bx$ can not have $\bw$ as substrings at positions $j$ and $k$ simultaneously since $\bu$ is $M$-RLL. In other words, either exactly one of $\bx_{j,\abs{\bw}}$ and $\bx_{k,\abs{\bw}}$ equals $\bw$ or none of them does. So given $\br$, we can get an upper bound on the probability of $H_\bw\parenv{a,j}\cap H_\bw\parenv{a,k}$ by assuming they are independent, i.e.,
    \begin{multline}
    \Pr\parenv{H_\bw\parenv{a,j}\cap H_\bw\parenv{a,k}|\br} \\
    \le
    \Pr\parenv{H_\bw\parenv{a,j}|\br}\Pr\parenv{H_\bw\parenv{a,k}|\br}. \label{eq:ubdind}
\end{multline}
We prove \eqref{eq:ubdind} rigorously by Lemma~\ref{lem:eq:ubdind} at the end of this section.

Denote the Hamming distance between $\br_0$ and $10^M$ by $d_0$, $\br_1$ and $\bu_1$ by $d_1$, $\br_2$ and $\bu_2$ by $d_2$, $\br_2$ and $\bu_2'$ by $d_2'$, $\br_3$ and $\bu_3$ by $d_3$, and $\br_4$ and $10^M$ by $d_4$. Let $\bw_l=10^M\bu$ and $\bw_r=\bu10^M$. The probability of occurrences increases if we only consider substrings $\bw_l$ or $\bw_r$. We have
\begin{align*}
    &\Pr\parenv{H_\bw\parenv{a,j}|\br} \\
    &\le \Pr\parenv{H_
    {\bw_l}\parenv{a,j}|\br}\\
    &=\!1\!-\!\parenv{\!1\!-\delta^{d_0+d_1+d_2}(1-\delta)^{\abs{\bw}-M-1-\parenv{d_0+d_1+d_2}}}^{\ceil{\frac{B}{4A}}} \\
    &\le \frac{B}{A}\delta^{d_0+d_1+d_2}(1-\delta)^{\abs{\bw}-M-1-\parenv{d_0+d_1+d_2}},
\end{align*}
and
\begin{align*}
    &\Pr\parenv{H_\bw\parenv{a,k}|\br} \\
    &\le \Pr\parenv{H_{\bw_r}\parenv{a,k+M+1}|\br}\\
    &= \!1\!-\!\parenv{\!1\!-\delta^{d_2'+d_3+d_4}(1-\delta)^{\abs{\bw}-M-1-\parenv{d_2'+d_3+d_4}}}^{\ceil{\frac{B}{4A}}} \\
    &\le \frac{B}{A}\delta^{d_2'+d_3+d_4}(1-\delta)^{\abs{\bw}-M-1-\parenv{d_2'+d_3+d_4}}.
\end{align*}
It follows from \eqref{eq:ubdind} that $\Pr\parenv{H_\bw\parenv{a,j}\cap H_\bw\parenv{a,k}}$ is less than or equal to
\begin{align*}
  &\quad \sum_{\br\in\Sigma^{\abs{\bw}}}\Pr\parenv{\br}\Pr\parenv{H_\bw\parenv{a,j}|\br}\Pr\parenv{H_\bw\parenv{a,k}|\br}\\
    & = \parenv{\frac{B}{A}}^2 \cdot\parenv{\sum_{\br_0\in\Sigma^{M+1}}\frac{1}{2^{\abs{\br_0}}}\delta^{d_0}(1-\delta)^{\abs{\br_0}-d_0}}\\
    &\quad \cdot\parenv{\sum_{\br_1\in\Sigma^{k-j}}\frac{1}{2^{\abs{\br_1}}}\delta^{d_1}(1-\delta)^{\abs{\br_1}-d_1}}\\
  &\quad \cdot\parenv{\sum_{\br_3\in\Sigma^{k-j}}\frac{1}{2^{\abs{\br_3}}}\delta^{d_3}(1-\delta)^{\abs{\br_3}-d_3}}\\
  &\quad\cdot\parenv{\sum_{\br_4\in\Sigma^{M+1}}\frac{1}{2^{\abs{\br_4}}}\delta^{d_4}(1-\delta)^{\abs{\br_4}-d_4}}\\
   &\quad \cdot\parenv{\sum_{\br_2\in\Sigma^{\abs{\bu}-(k-j)}}\frac{1}{2^{\abs{\br_2}}}\delta^{d_2+d_2'}(1-\delta)^{2\abs{\br_2}-(d_2+d_2')}}\\
     &= \parenv{\frac{B}{A}}^2\cdot \frac{1}{2^{2M+2+2(k-j)}}\\
    & \quad \cdot\parenv{\sum_{\br_2\in\Sigma^{\abs{\bu}-(k-j)}}\frac{1}{2^{\abs{\br_2}}}\delta^{d_2+d_2'}(1-\delta)^{2\abs{\br_2}-(d_2+d_2')}}.
\end{align*}
Let $d^\circ$ denote the Hamming distance between $\bu_2$ and $\bu_2'$. Among the $\abs{\bu_2}-d^\circ$ positions where $\bu_2$ and $\bu_2'$ are the same, suppose $\bu_2$ differs from $\br_2$ in $v$ of them. Among the $d^\circ$ positions where $\bu_2$ differs from $\bu_2'$, suppose $\bu_2$ differs from $\br_2$ in $t$ of them.  It follows that $d_2 = v+t$ and $d_2'=d^\circ+v-t$. Thus, we further have 
\begin{align*}
    &\sum_{\br_2\in\Sigma^{\abs{\bu}-(k-j)}}\frac{1}{2^{\abs{\br_2}}}\delta^{d_2+d_2'}(1-\delta)^{2\abs{\br_2}-(d_2+d_2')} \\
    =& \sum_{\br_2\in\Sigma^{\abs{\bu}-(k-j)}}\frac{1}{2^{\abs{\br_2}}} \delta^{2v+d^\circ}(1-\delta)^{2\abs{\br_2}-2v-d^\circ}\\
    =& \sum_{v=0}^{\abs{\br_2}-d^\circ}{\abs{\br_2}-d^\circ\choose v} \frac{2^{d^{\circ}}}{2^{\abs{\br_2}}}\delta^{2v+d^\circ}(1-\delta)^{2\abs{\br_2}-2v-d^\circ} \\
    =& \frac{\parenv{2\delta(1-\delta)}^{d^\circ}}{2^{\abs{\br_2}}} \sum_{v=0}^{\abs{\br_2}-d^\circ} {\abs{\br_2}-d^\circ\choose v} \\
    &\quad\cdot\parenv{\delta^2}^v \parenv{(1-\delta)^2}^{(\abs{\br_2}-d^\circ)-v}\\
    =& \frac{1}{2^{\abs{\br_2}}} \parenv{2\delta(1-\delta)}^{d^\circ} \parenv{\delta^2+(1-\delta)^2}^{\abs{\br_2}-d^\circ}.
\end{align*}
Since $\abs{\br_2} = \abs{\bw} - (k-j)$,
\begin{align*}
    &\quad\Pr\parenv{H_\bw\parenv{a,j}\cap H_\bw\parenv{a,k}} \\
    &\le \parenv{\frac{B}{A}}^2\frac{\parenv{2\delta(1-\delta)}^{d^\circ}\parenv{\delta^2+(1-\delta)^2}^{\abs{\br_2}-d^{\circ}}}{2^{\abs{\bw}+(k-j)}}.
\end{align*}

Note that $\bu_2$ is the $\abs{\br_2}$-suffix of $\bu$ and $\bu_2'$ is the $\abs{\br_2}$ prefix of $\bu$. With $\abs{\bu}=n$, the number of $n$-strings whose $\abs{\br_2}$-suffix and $\abs{\br_2}$-prefix are at Hamming distance $d^\circ$ is $2^{n-\abs{\br_2}}{\abs{\br_2}\choose d^\circ}$ since an $n$-string can be uniquely determined by its $\abs{\br_2}$-prefix and the mismatches. Therefore, 
\begin{align}
    &\sum_{\substack{\bw:\bw=10^M\bu10^M\\ \bu\in R_M^n}}\;\sum_{1\le\abs{j-k}<\abs{\bu}}\Pr\parenv{H_\bw\parenv{a,j}\cap H_\bw\parenv{a,k}} \nonumber\\
    &\le \sum_{1\le\abs{j-k}<\abs{\bu}}\;\sum_{\substack{\bw:\bw=10^M\bu10^M\\ \bu\in \Sigma^n}}\Pr\parenv{H_\bw\parenv{a,j}\cap H_\bw\parenv{a,k}}\nonumber\\
    & \le L\abs{\bu}\cdot\sum_{d^\circ =0}^{\abs{\br_2}} 2^{\abs{\br_1}}{\abs{\br_2}\choose d^\circ}\cdot \parenv{\frac{B}{A}}^2\nonumber\\
    &\quad \cdot\frac{\parenv{2\delta(1-\delta)}^{d^\circ}\parenv{\delta^2+(1-\delta)^2}^{\abs{\br_2}-d^\circ}}{2^{\abs{\bw}+k-j}} \nonumber\\
    & = \parenv{\frac{B}{A}}^2\frac{Ln}{2^{\abs{\bw}}}.\label{eq:wapplbdin32}
\end{align}

    \item $\abs{\bu}\le \abs{j-k}<\abs{\bw}$. The two occurrences of $\bw$ in $\bx_1$ and $\bx_2$ are plotted in Figure~\ref{fig:twooccurwcase3}. 
   \begin{figure}[h]
	\begin{center}
		\scalebox{0.65}{\begin{tikzpicture}[
		node distance = 0cm,
		seq/.style={minimum width=.3cm, minimum height=0.5cm, shape=rectangle,draw,font=\small,inner sep=0mm},
		marker/.style={pattern=dots, pattern color=black},
		nonoverlap/.style={pattern = north east lines, pattern color = green},
		overlap/.style={pattern=north west lines, pattern color=red},
		mu/.style={},
		]
		
		\node (root) [seq,minimum width=12cm] {};
		\node (rootname) [left = of root, xshift =-.3cm,seq, draw=none] {$\by_a$};

		\node (x1) [below = of root,yshift=-.5cm, seq,minimum width=12cm] {};
		\node (x1name) [left = of x1, xshift = -.3cm, seq, draw=none] {$\bx_1$};
		\node (w1lmk) [right = of x1name,xshift = 1cm,seq,minimum width = 0.8 cm]{$10^M$};
		\node (x1u) [right = of w1lmk, seq, overlap, minimum width = 3.4cm]{$\bu$};
		\node (w1rmk) [right = of x1u, seq, minimum width = 0.8cm]{$10^M$};

		\node (x2) [below = of x1,yshift=-.5cm, seq,minimum width=12cm] {};	
		\node (x2name) [left = of x2,xshift =-.3cm]{$\bx_2$};
        \node (w2lmk) [right = of x2name, xshift = 5.7cm, seq, minimum width =0.8cm]{$10^M$};
        \node (x2u) [right = of w2lmk, seq, overlap, minimum width = 3.4cm]{$\bu$};
        \node (w2rmk) [right = of x2u, seq, minimum width =0.8cm]{$10^M$};
        \draw [dashed] ([xshift = -0.4cm, yshift = 2cm]w1lmk.south) -- ([xshift = -0.4cm,yshift = 0cm]w1lmk.south);
        \node [above = of w1lmk,xshift = -0.4cm,yshift = 1.5cm ]{$j$};
        
        \draw [dashed] ([xshift = -0.4cm, yshift = 3cm]w2lmk.south) -- ([xshift = -0.4cm,yshift = 0cm]w2lmk.south);
        \node [above = of w2lmk,xshift = -0.4cm,yshift = 2.5cm ]{$k$};
        \draw[decorate,decoration={brace,amplitude=5pt,raise=1pt},yshift=-1pt]   ([xshift = -0.4cm]w1lmk.north)--([xshift = 0.4cm]w1rmk.north) node [black,midway, yshift = 0.35cm] 
    {$\bw$};
    \draw[decorate,decoration={brace,amplitude=5pt,raise=1pt},yshift=-1pt]   ([xshift = -0.4cm]w2lmk.north)--([xshift = 0.4cm]w2rmk.north) node [black,midway, yshift = 0.35cm] 
    {$\bw$};
		\end{tikzpicture}}
	\end{center}
	\caption{Relative position of the two occurrences of $\bw$ at position $j$ and $k$ when $\abs{\bu}\le \abs{j-k}<\abs{\bw}$.}
	\label{fig:twooccurwcase3}
\end{figure}

It can be seen that the prefix $10^M\bu$ of $\bw$ in $\bx_1$ and $\bw$ in $\bx_2$ are descendants of non-overlapping substrings of $\by_a$ and thus independent. We can write
\begin{align*}
    &\Pr\parenv{H_\bw\parenv{a,j}\cap H_\bw\parenv{a,k}} \\
    &\le \Pr\parenv{H_{\bw_l}\parenv{a,j}\cap H_\bw\parenv{a,k}}\\
    & = \Pr\parenv{H_{\bw_l}\parenv{a,j}}\Pr\parenv{H_\bw\parenv{a,k}}\\
    &\le \frac{\cS_{\delta}\parenv{\abs{\bw_l},\ceil{\frac{B}{4A}}}\cS_{\delta}\parenv{\abs{\bw},\ceil{\frac{B}{4A}}}}{2^{\abs{\bw_l}+\abs{\bw}}}\\
    &\le \parenv{\frac{B}{A}}^2 \frac{1}{2^{2\abs{\bw}-M-1}}.
\end{align*}
It follows that 
\begin{align}
     &\sum_{\abs{\bu}\le\abs{j-k}< \abs{\bw}}\Pr\parenv{H_\bw\parenv{a,j}\cap H_\bw\parenv{a,k}}\nonumber \\
     &\le \parenv{\frac{B}{A}}^2\frac{L\cdot 2(M+1)}{2^{2\abs{\bw}-M-1}}. \label{eq:wapplbdin33}
\end{align}
\end{itemize}

Thus, combining \eqref{eq:wapplbdin31}, \eqref{eq:wapplbdin32}, \eqref{eq:wapplbdin33} gives 
\begin{align*}
     &\sum_{\substack{\bw:\bw=10^M\bu10^M\\ \bu\in R_M^n}}\;\sum_{\substack{1\le j, k \le \ceil{L/2}-\abs{\bw}\\j\ne k}} \Pr\parenv{H_\bw\parenv{a,j}\cap H_\bw\parenv{a,k}} \\
     &\le  \sum_{\substack{\bw:\bw=10^M\bu 10^M\\\bu\in R_M^n}}\left(\sum_{\abs{j-k}\ge\abs{\bw}} \Pr\parenv{H_\bw\parenv{a,j}\cap H_\bw\parenv{a,k}} \right. \\
     & \quad +
    \sum_{1\le\abs{j-k}<\abs{\bu}} \Pr\parenv{H_\bw\parenv{a,j}\cap H_\bw\parenv{a,k}}\\
     &\quad \left.+\sum_{\abs{\bu}\le \abs{j-k}\le\abs{\bw}}\Pr\parenv{H_\bw\parenv{a,j}\cap H_\bw\parenv{a,k}} \right)\\
     &\le\frac{L^2B^2}{A^2\cdot 2^{2\abs{\bw}}}\cdot\abs{R_M^n}
     + \frac{B^2Ln}{A^2\cdot 2^{\abs{\bw}}}+ \frac{2B^2L(M+1)}{A^2\cdot 2^{2\abs{\bw}-M-1}}\cdot \abs{R_M^n}\\
    &\le\frac{B^2L^2}{A^22^{\abs{\bw}}}\parenv{1 + \frac{n+M+1}{L} }.     
\end{align*}
\end{IEEEproof}

We present a lemma from which inequality \eqref{eq:ubdind} follows directly.
\begin{lem}\label{lem:eq:ubdind}
Let $\br$ be any string of length $n$ with $m$ iid $\delta$-edit descendants. For a string $\bv,\abs{\bv}<n$ and $1\le j< k\le n-\abs{\bv}+1$, let $\cJ(\bv),\cK(\bv)$ denote the events that there exists a descendant of $\br$ whose $j$-th, $k$-th $\abs{\bv}$-substring equal $\bv$, respectively. We have 
\begin{align*}
    \Pr\parenv{\cJ\parenv{\bv}\cap\cK\parenv{\bv}} \le \Pr\parenv{\cJ\parenv{\bv}}\Pr\parenv{\cK\parenv{\bv}},
\end{align*}
if the $\parenv{\abs{\bv}-(k-j)}$-suffix and $\parenv{\abs{\bv}-(k-j)}$-prefix of $\bv$ are not the same.
\end{lem}
\begin{IEEEproof}
If the $\parenv{\abs{\bv}-(k-j)}$-suffix and $\parenv{\abs{\bv}-(k-j)}$-prefix of $\bv$ are not the same, then in any descendant $\bx$, $\bv$ can not be both the $j$-th and the $k$-th substring. Therefore, in $\bx$, exactly one of the following three mutually exclusive events holds: i) $\bx_{j,\abs{\bv}} = \bv$, ii) $\bx_{k,\abs{\bv}}=\bv$, iii) $\bx_{j,\abs{\bv}} \ne \bv$ and $\bx_{k,\abs{\bv}}\ne\bv$. Let $p_j$ denote the probability of $\bx_{j,\abs{\bv}} = \bv$ and $p_k$ denote the probability of $\bx_{k,\abs{\bv}}=\bv$. We have 
\begin{align*}
    \Pr\parenv{ \bx_{j,\abs{\bv}} \ne \bv\cap \bx_{k,\abs{\bv}}\ne\bv} = 1-p_j-p_k.
\end{align*}
Therefore, among the $m$ iid descendants of $\br$,
\begin{align*}
    &\quad \Pr\parenv{ \cJ\parenv{\bv}\cap\cK\parenv{\bv}} \\
    &= \Pr\parenv{\cJ(\bv)} + \Pr\parenv{\cK(\bv)} + \Pr\parenv{\bar\cJ(\bv)\cap\bar\cK(\bv)} -1 \\
    &= \parenv{1-\parenv{1-p_j}^m} + \parenv{1-\parenv{1-p_k}^m} \\
    &\quad + \parenv{1-p_j-p_k}^m -1\\
    &=1-\parenv{1-p_j}^m - \parenv{1-p_k}^m + \parenv{1-p_j-p_k}^m.
\end{align*}
On the other hand, 
\begin{align*}
    &\quad \Pr\parenv{\cJ\parenv{\bv}}\Pr\parenv{\cK\parenv{\bv}} \\
    &= \parenv{1-\parenv{1-p_j}^m}\parenv{1-\parenv{1-p_k}^m}\\
    & = 1-\parenv{1-p_j}^m-\parenv{1-p_k}^m \\
    &\quad + \parenv{1-p_j}^m\parenv{1-p_k}^m.
\end{align*}
The desired inequality thus follows by noting that $1-p_j-p_k\le \parenv{1-p_j}\parenv{1-p_k}$.
\end{IEEEproof}
Inequality \eqref{eq:ubdind} can be obtained by replacing $\cJ(\bv)$ and $\cK(\bv)$ with $H_\bw\parenv{a,j}$ and $H_\bw\parenv{a,k}$, respectively.

\section{Proofs of Lemma~\ref{lem:numlck2hf} and Lemma~\ref{lem:idsstrins}}\label{app:numlck2hf}\label{app:insstrins}
\numlcktohf*
\begin{IEEEproof}
Equally parse each of $Y_{\ceil{B/2}+1},\ldots,Y_B$ into segments of length $2^{M+7}$. So that every $Y_b$ contains $\floor{\frac{\abs{Y_b}}{2^{M+7}}}$ segments. We show that among these $\sum_{b=\floor{\frac{B}{2}}+1}^B\floor{\frac{\abs{Y_b}}{2^{M+7}}}$ segments, a constant fraction of them contain a chunk of length over $2^{M-4}$.

Pick an arbitrary segment, denoted $\bz$. Consider the two halves of $\bz$. The second half of $\bz$, which is of length $2^{M+6}$, is by itself a Bernoulli(1/2) process going forward. We study the first time a run of $M$ 0's appears in this process. By the union bound, with probability at least $1-\frac{2^{M-5}}{2^{M}}$, there exist no runs of $M$ 0s in the first $2^{M-5}$ bits. Moreover, the average position of the end of the first run of $M$ 0s in a Bernoulli(1/2) process is $2^{M+1}-2$~\cite{sedgewick2013introduction}. Therefore, by Markov's inequality, with probability at least $1-\frac{2^{M+1}-2}{2^{M+6}}$, there is a $0^M$ within the first $2^{M+6}$ bits. So the first time we see $0^M$ is after $2^{M-5}$ bits and before $2^{M+6}$ bits (i.e., the first $0^M$ is within the last $2^{M+6}-2^{M-5}$ bits) with probability at least 
\[1-\frac{2^{M-5}}{2^{M}}-\frac{2^{M+1}-2}{2^{M+6}} \ge 1-\frac{1}{2^4}.\]
Similarly, the first half of $\bz$ can be regarded as a reversed Bernoulli(1/2) process. So we also have with probability at least $1-\frac{1}{2^4}$, the first $0^M$ (counting backwards) is within the first $2^{M+6}-2^{M-5}$ bits. Clearly, a chunk exists between these two occurrences of $0^M$. So with probability at least $1-\frac{1}{2^3}$, $\bz$ contains a chunk of length at least $2^{M-4}$. Since this property holds for all such segments of length $2^{M+7}$, by the Markov inequality, with probability at least $1-\frac{1}{6}$, at least $\frac{1}{4}$ of the segments in $Y_{\ceil{B/2}+1}\cdots Y_B$ contain a chunk of length at least $2^{M-4}$. The desired result is derived by noting $\abs{Y_b}\ge L/2$.
\end{IEEEproof}

\lemidsstrins*
\begin{IEEEproof}
We compute $\Pr\parenv{(Y_{b_1})_{i_1,h} = (Y_{b_2})_{i_2,h}}$ as $(b_1,b_2),(i_1,i_2)$ take different values in the following three cases:
\begin{itemize}
    \item $J_{b_1}\ne J_{b_2}$ or $\abs{i_1-i_2} \ge h$. If $J_{b_1}\ne J_{b_2}$, then $Y_{b_1}$ and $Y_{b_2}$ have different ancestors and are thus independent. It follows that their substrings are also independent. If $\abs{i_1-i_2}\ge h$, then $(Y_{b_1})_{i_1,h}$ and $(Y_{b_2})_{i_2,h}$ are descendants of non-overlapping substrings of the source alphabet and are thus also independent. The desired result follows from the fact that $(Y_{b_1})_{i_1,h}$ and $(Y_{b_2})_{i_2,h}$ are both Bernoulli(1/2) processes by themselves.
    \item $b_1=b_2,\abs{i_1-i_2}<h$. In this case, $(Y_{b_1})_{i_1,h}$ and $(Y_{b_2})_{i_2,h}$ are overlapping substrings of a single source block. Again, $Y_{b_1}$ is Bernoulli(1/2) by itself. So the probability of $(Y_{b_1})_{i_1,h} = (Y_{b_2})_{i_2,h}$ is the same as that when $(Y_{b_1})_{i_1,h}$ and $(Y_{b_2})_{i_2,h}$ are independent.
    \item $J_{b_1}=J_{b_2}, b_1\ne b_2, \abs{i_1-i_2}<h$. Let $J_{b_1}=J_{b_2}=a$. Assume $i_1< i_2$ without loss of generality. In this case, $(Y_{b_1})_{i_1,h}$ and $(Y_{b_2})_{i_2,h}$ are two independent $\delta$-edit descendants of $(\gx_a)_{i_1,h}$ and $(\gx_a)_{i_2,h}$, respectively. So $\Pr\parenv{(Y_{b_1})_{i_1,h} = (Y_{b_2})_{i_2,h}}$ is uniquely determined by the Hamming distance between $(\gx_a)_{i_1,h}$ and $(\gx_a)_{i_2,h}$. Moreover, the distribution of the Hamming distance between $(\gx_a)_{i_1,h}$ and $(\gx_a)_{i_2,h}$ is the same as the distribution of the Hamming distance between two independent Bernoulli(1/2) process of length $h$. Therefore, we can assume $(Y_{b_1})_{i_1,h}$ and $(Y_{b_2})_{i_2,h}$ are independent and thus $\Pr\parenv{ (Y_{b_1})_{i_1,h}=(Y_{b_2})_{i_2,h}}=\frac{1}{2^h}$.
\end{itemize}
\end{IEEEproof}

\section{Summations}\label{app:summation}
For integers $ b\ge a  $ and $\beta>1$, summations of the forms $\sum_{n = a}^b \parenv{1-\frac{1}{\beta}}^n $ and $ \sum_{n=a}^b n\parenv{1-\frac{1}{\beta}}^n$ appear in the proofs of Theorem~\ref{thm:vllbdlargeM} and Theorem~\ref{thm:vlubd}. Let $x = 1-\frac{1}{\beta}$. The limits of these sums in a certain asymptotic regime is discussed bolew.
\subsection{Asymptotic behavior of $\sum_{n=a}^b x^n$}
We have 
\begin{align*}
    \sum_{n=a}^b x^n &= \frac{x^a \parenv{1-x^{b-a+1}}}{1-x} \\
    &= \beta\parenv{1-\frac{1}{\beta}}^a \parenv{1-\parenv{1-\frac{1}{\beta}}^{b-a+1}}.
\end{align*}
If $b-a = \omega(\beta)$, then as $\beta\rightarrow\infty$,
\begin{align*}
    \parenv{1-\frac{1}{\beta}}^{b-a+1} = \parenv{\parenv{1-\frac{1}{\beta}}^{\beta}}^{\frac{b-a+1}{\beta}} = o(1).
\end{align*}
It follows that
\begin{align}
    \sum_{n=a}^b\parenv{1-\frac{1}{\beta}}^n &=\beta\parenv{1-\frac{1}{\beta}}^a \parenv{1+o(1)}\nonumber\\
    &=\beta e^{-\nicefrac{a}{\beta}}\parenv{1+o(1)}. \label{eq:sumbl}
\end{align}
\subsection{Asymptotic behavior of $\sum_{n=a}^b nx^n$} \label{app:sum2}
We have
\begin{align*}
    &\quad \sum_{n=a}^b n x^n = x \sum_{n=a}^b n x^{n-1} = x\parenv{\sum_{n=a}^b x^n}' \\
    &= x\parenv{ \frac{x^a \parenv{1-x^{b-a+1}}}{1-x}}' \\
    &= x\frac{\parenv{ax^{a-1}-(b+1)x^b}\parenv{1-x}+\parenv{x^a-x^{b+1}}}{\parenv{1-x}^2} \\
    & = \beta^2\parenv{\!\parenv{\frac{a-1}{\beta}+1}\parenv{1-\frac{1}{\beta}}^a  + \parenv{\frac{b}{\beta}+1} \parenv{1-\frac{1}{\beta}}^{b+1}}.
\end{align*}
If $\frac{b}{\beta} = \omega\parenv{1}$, then as $\beta\rightarrow \infty$,
\begin{align*}
    \parenv{\frac{b}{\beta}+1} \parenv{1-\frac{1}{\beta}}^{b+1} &= \parenv{\frac{b}{\beta}+1} \parenv{\parenv{1-\frac{1}{\beta}}^\beta}^{\frac{b+1}{\beta}}\\
    &=o(1).
\end{align*}
It follows that 
\begin{align}
    \sum_{n=a}^b n \parenv{1-\frac{1}{\beta}}^n &= \beta^2\parenv{\parenv{\frac{a-1}{\beta}+1}\parenv{1-\frac{1}{\beta}}^a  + o(1)}\nonumber\\
    &=\beta^2\parenv{\frac{a-1}{\beta}+1}e^{-\nicefrac{a}{\beta}}\parenv{1+o(1)}. \label{eq:sumlbl}
\end{align}
\printbibliography

\begin{IEEEbiographynophoto}{Hao Lou} 
(S'18) is a PhD candidate in the Department of Electrical and Computer Engineering at the University of Virginia. His research interests include data deduplication, stochastic and information-theoretic modeling of DNA mutations, compression of metagenomic sequencing data, computational biology and machine learning. He received his Bachelor's degree from Xi'an Jiaotong University, China in 2017.
\end{IEEEbiographynophoto}

\begin{IEEEbiographynophoto}{Farzad Farnoud (Hassanzadeh)}
(M'13) is an Assistant Professor in the Department of Electrical and Computer Engineering and the Department of Computer Science at the University of Virginia. Previously, he was a postdoctoral scholar at the California Institute of Technology. 

He received his MS degree in Electrical and Computer Engineering from the University of Toronto in 2008. From the University of Illinois at Urbana-Champaign, he received his MS degree in mathematics and his Ph.D. in Electrical and Computer Engineering in 2012 and 2013, respectively. His research interests include coding for storage, data compression, probabilistic modeling and analysis, and machine learning. He is the recipient of the 2013 Robert T. Chien Memorial Award from the University of Illinois for demonstrating excellence in research in electrical engineering, the 2014 IEEE Data Storage Best Student Paper Award, and a 2022 NSF CAREER Award.
\end{IEEEbiographynophoto}
\end{document}